\documentstyle[aaspp4]{article}

\received{}
\accepted{}
\journalid{VOL}{JOURNAL DATE}
\articleid{START PAGE}{END PAGE}
\paperid{MANUSCRIPT ID}

\begin{document}

%\magnification=1200
%\def\pmb#1{\setbox0=\hbox{$#1$}
% \kern-.025em\copy0\kern-\wd0
% \kern-.05em\copy0\kern-\wd0
% \kern-.025em\raise.0433em\box0 }
\def\gtsim {>\kern-1.2em\lower1.1ex\hbox{$\sim$}}
\def\ltsim {<\kern-1.2em\lower1.1ex\hbox{$\sim$}}
\def\ref{\hangindent=1pc \hangafter=1 \noindent}
\def\pmb#1{\mbox{\boldmath$#1$}}

\def\gtsim {>\kern-1.2em\lower1.1ex\hbox{$\sim$}}
\def\ltsim {<\kern-1.2em\lower1.1ex\hbox{$\sim$}}
\def\ref{\hangindent=1pc \hangafter=1 \noindent}

\title{\bf Surface $r$-modes and burst oscillations of neutron stars}
\author{Umin Lee}
\affil{Astronomical Institute, Tohoku University, Sendai, Miyagi 980-8578, Japan
\\ lee@astr.tohoku.ac.jp}

\begin{abstract} We study the $r$-modes propagating in steadily mass accreting,
nuclear burning, and geometrically thin envelopes on the surface of rotating neutron stars.
For the modal analysis, we construct the envelope models which are fully radiaitive
or have a convective region.
We simply call the former radiative models and the latter convective models in this paper.
As the angular rotation frequency $\Omega$ is increased,
the oscillation frequency $\omega$ of the $r$-modes in the thin envelopes
deviates appreciably from the asymptotic frequency $\omega=2m\Omega/l^\prime(l^\prime+1)$
defined in the limit of $\Omega\rightarrow 0$, where $\omega$ is the frequency observed 
in the corotating frame of the star, and $m$ and $l^\prime$ are 
the indices of the spherical harmonic function
$Y_{l^\prime}^m$ representing the angular dependence of the modes.
We find that the amplitudes of the fundamental $r$-modes with no radial nodes of the
eigenfunctions are
strongly confined to the equatorial region, and 
$\omega$ becomes only weakly dependent on $\Omega$, 
gathering in a frequency range of $\omega/2\pi\ltsim 10$Hz, at rapid rotation rates.
We also find that the fundamental $r$-modes in the convective models
are destabilized by strong nuclear burning in the convective region.
Because of excessive heating by nuclear buring,
the corotating-frame oscillation frequency $\omega$ of the $r$-modes in the convective models 
becomes larger, and hence the inertial-frame oscillation frequency $|\sigma|$
becomes smaller, than those
of the corresopnding $r$-modes in the radiative models, where $\sigma=\omega-m\Omega$
is negative for the $r$-modes of positive $m$.
We find that the relative frequency change
$f=-(\sigma_{conv}-\sigma_{rad})/\sigma_{rad}$ is always positive and becomes less than $\sim$0.01
for the fundamental $r$-modes of $l^\prime>|m|+1$ at $|\sigma_{rad}|/2\pi\sim$300Hz for $m=1$
or at $|\sigma_{rad}|/2\pi\sim$600Hz for $m=2$, and that
we need to consider the $r$-modes of $l^\prime$
much larger than $|m|$ for values of $f$ as small as $\sim$0.001, where
$\sigma_{conv}$ and $\sigma_{rad}$ denote the 
oscillation frequencies for the
convective and the radiative envelope models, respectively.

\end{abstract}

\keywords{instabilities --- stars: neutron --- stars: oscillations --- stars : rotation}

\section{Introduction}

X-ray flux fluctuations observed in type I X-ray bursts in LMXB's 
(see, e.g., Strohmayer et al 1996, 1997 for so called burst oscillations)
are quasi-periodic oscillations whose frequencies are found in a frequency
range around $\sim$300Hz or around $\sim$600Hz.
During a burst, from the onset to the tail,
the oscillation frequency in a given source usually increases with time by a small amount 
and gets saturated to an asymptotic frequency in the tail.
The observed frequency drifts are generally less than $\sim$1\% of the mean frequency
(e.g., Cumming \& Bildsten 2000).
However, the time evolution of burst oscillations are not always that simple.
In fact, some X-ray burst sources have been reported to show a temporal decrease
of the oscillation frequency in the tail of bursts
(e.g., Strohmayer 1999, Miller 2000, Muno et al. 2000), or to show
simultaneously two oscillation frequencies separated by a few Hz in a burst
(Miller 2000, Galloway et al 2001, Muno et al 2002).
Burst oscillations could be very important for neutron star physics. 
On the assumption that the oscillation frequencies are directly tied to the spin
frequency of the star,
the stability of the asymptotic frequency found in the tail 
of bursts separated by a few years has
been used to determine some of the physical parameters of binary systems 
(e.g., Strohmayer et al 1998, Gilles et al 2002).
For recent observational and theoretical developements concerning burst oscillations as well as 
thermonuclear bursts on neutron stars,
see a review paper by Strohmayer \& Bildsten (2003).

As a model for burst oscillations, 
Strohmayer et al (1997) proposed that the oscillations are
produced by inhomogeneous hot burning spots, the burning fronts of which are expanding
laterally on the surface of a rotating neutron star, and that
the frequency drifts are associated with the vertical expansion and contraction of the 
burning spots conserving angular momentum.
The model was later refined by
Cumming \& Bildsten (2000), and Cumming et al (2001) concluded that the possible amount of  
relative frequency changes expected in the model was by a factor of 2 to 3 smaller 
than the largest observed values (see also Muno, \"Ozel, \& Chakrabarty 2002).
This model also has a difficulty in explaining the persistence of the oscillations to the
tail of a burst, in which no inhomogeneity to produce the oscillations
will remain because all the surface area has been reached by the burning fronts.
Spitkovsky, Levin, \& Ushomirsky (2002), on the other hand,
discussed the global hydrodynamical flow in the ocean of an accreting neutron star,
by taking into account the rapid rotation and the lift-up of the burning ocean, and
conjectured that Jupiter-type vortices created in strong zonal fluid currents driven 
in a nuclear burst are responsible for burst oscillations in the tail.

It was Heyl (2001) who suggested
that the burst oscillations could be produced by the $r$-modes 
of low azimuthal wavenumber $m=1$ or $m=2$ traveling in the surface
ocean of the neutron star.
The $r$-modes are retrograde toroidal modes
and are called Rossby waves in geophysical context
(see, e.g., Greenspan 1968, Pedlosky 1987).
The $r$-modes in mass-accreting envelopes on the surface of a 
$slowly$ rotating neuton star have been discussed by Strohmayer \& Lee (1996), who
found that
the fundamental $r$-modes with no radial nodes of the eigenfunctions
are all pulsationally stable except for the cases of mass-acceretion rates $\dot M$
as low as $\dot M/\dot M_{Edd}\sim0.001$, where $\dot M_{Edd}$ denotes the Eddington rate.
In this paper, we extend the previous analysis to the cases of rapidly
rotating neutron stars, considering that the spin frequencies $\Omega$
of many neutron stars in LMXB's are now believed to be in a rather narrow range around 
$\Omega/2\pi\sim300$Hz.
The main points behind the $r$-mode model for burst oscillations
may be described as follows.
Let us assume that strong nuclear burning in the convective phase of a nuclear burst excites 
surface $r$-modes whose time and azimuthal dependence is given by 
$\exp(i[\omega_{conv}t+m\phi])$, where $\omega$ denotes the oscillation frequency in the 
corotating frame and the integer $m$ is the azimuthal wave number.
For the $r$-modes, $\omega$ is proportional to and is much smaller than the spin 
frequency $\Omega$ of the star, and $\omega$ is positive for positive $m$
in the convention we use in this paper.
Observers measure the inertial frame frequency $\sigma_{conv}\equiv\omega_{conv}-m\Omega$,
the magnitude of which is $|\sigma_{conv}|=m\Omega-\omega_{conv}$.
At the end of the burst, the ocean cools and becomes radiative, and the rotating frame
oscillation frequency decreases to $\omega_{rad}$ and the magnitude of 
the inertial frame frequency increases to $|\sigma_{rad}|=m\Omega-\omega_{rad}$.
This model then predicts a fractional increase in the measured inertial frame frequency
from the start of the burst to the asymptotic end, which may be given by
$f=-(\sigma_{conv}-\sigma_{rad})/\sigma_{rad}$.
To estimate possible magnitudes of the quantity $f$, we calculate in this paper
the $r$-modes for radiative envelopes and convective envelopes, assuming that the envelopes are
in steady state.
\S 2 and 3 briefly describe method of calculation employed in this paper, and
\S 4 is for numerical results, and \S 5 and 6 are for discussion and conclusion.

\section{Envelope Calculation}

Method of calculation for steadily mass-accreting, nuclear-burning, thin 
surface envelopes 
is almost the same as that given in Strohmayer \& Lee (1996).
No effects of general relativity are included in this paper.
If we use the independent variable defined as
$$
z=M-M_r, \eqno (1)
$$ 
where $M$ and $M_r$ are the total mass of the star and the mass contained
in the sphere of radius $r$,
the basic ordinary differential equations for mass-accreting and nuclear-burning thin
envelopes in steady state are 
$$
{dr\over dz}=-{1\over 4\pi r^2\rho}, \eqno (2)
$$
$$
{dp\over dz}={GM_r\over 4\pi r^4}, \eqno (3)
$$
$$
{dL_r\over dz}=-\epsilon_N+\epsilon_\nu
+{\dot M}\left({dU\over dz}-{p\over\rho^2}{d\rho\over dz}\right), \eqno (4)
$$
and the energy transfer equation in convection zones is
$$
{dT\over dz}={GM_r\over 4\pi r^4}{T\over p}
\left({\partial\ln T\over\partial\ln p}\right)_{ad}
\equiv{GM_r\over 4\pi r^4}{T\over p}\nabla_{ad}, \eqno(5)
$$
and that in radiative zones is
$$
{dT\over dz}={3\kappa L_{r}\over 64\pi^2acr^4T^3}, \eqno (6)
$$
where $\rho$ is the mass density, $p$ is the pressure,
$T$ is the temperature, $U$ is the internal energy per gram,
$\kappa$ is the opacity,
$\epsilon_N$ is the nuclear energy generation rate
per gram, $\epsilon_\nu$ is the neutrino energy loss rate per gram,
$L_r$ is the luminosity at $r$,
$\dot M=-4\pi r^2\rho v_r$ is the mass accretion rate, and 
$G$, $a$, and $c$ are the gravitational constant,
the radiation constant, and the velocity of light, respectively.
The structure of the chemical composition, $X_i$, in steady state is governed by
$$
{dX_j\over dz}={m_j\over \rho\dot M}\sum_i{n_{ji}r_i}, \eqno (7)
$$
where $m_j$ is the mass of the particle $j$,
$r_i$ is the rate of the reaction $i$, and $n_{ji}$ is the number of particles $j$
created ($n_{ji}>0$) or destroyed ($n_{ji}<0$) in a reaction $i$.
The equation of state $p=p(\rho,T,X_j)$ is that of a perfect gas with radiation,
and is given by $p=p_I+p_e+p_r$, where $p_I=n_Ik_BT$ with $n_I$ and $k_B$
being the number density of ions and the Boltzmann constant, 
$p_e$ is the electron gas pressure, and $p_r=aT^4/3$ is the radiation pressure.
The electron gas can be arbitrarily degenerate and relativistic.
For hydrogen burning energy generation rates, we use the formulae given in Reeves (1965).
The reaction rates of the CNO cycle in high temperature environment
are limited by the $\beta$-decays of $^{14}O$ and $^{15}O$, leading to
the saturation of the rates to $\epsilon_{CNO}^{max}=5.9\times10^{15} Z_{CNO}$ 
(e.g., Wallace \& Woosley 1981),
where $Z_{CNO}=^{12}C+^{14}N+^{16}O$.
For helium burning we include the $3\alpha$ and $^{12}C(\alpha,\gamma)^{16}O$
reactions, the rates of which are given in Fowler, Caughlan, \& Zimmerman (1975).
The rate of the carbon burning $^{12}C+^{12}C$ reaction is 
from Patterson, Winkler, \& Zaidins (1969),
and Arnett \& Truran (1969).
We apply both weak and strong electron screening corrections to the
nuclear energy generation rates following Salpeter \& Van Horn (1969).
For the energy loss rate due to neutrino emission we employ formulae given by 
Beaudet, Petrosian, \& Salpeter (1967).
For opacity, we use a subroutine written by Paczy\'nski (1970) who adopted
Cox \& Stewart (1970) radiative opacity and the conductive opacity
by Hubbard \& Lampe (1969) and Canuto (1970).

For given $M$, $R$, and $\dot M$, 
we integrate the set of differential equations from the surface $z=0$ to
the bottom $z=z_b$ of the envelope.
At the surface of the envelope, where we assume $r=R$ and $M_r=M$, 
the surface tempetrature $T_s$ and pressure $p_s$ for a
surface luminosity $L_s$ are given by
$$
T_s=\left({L_{s}/ 4\pi R^2\sigma_{SB}}\right)^{1/4}, \eqno (8)
$$
and 
$$
p_s=g_s/\kappa(\rho_s,T_s,X_i), \eqno (9)
$$
where $g_s=GM/R^2$,  
and $\sigma_{SB}=ac/4$ is the Stefan-Boltzmann constant.
The surface luminosity $L_s$ is determined so that the inner boundary
condition $L_r=0$ is satisfied at $z=z_b$.
The envelope models calculated in this way are fully radiative with no convective regions,
and are simply called radiative models in this paper.

As discussed by Hanawa \& Sugimoto (1982), there developes a convective zone 
in the nuclear burning region of the surface envelope immediately after the onset of
a nuclear flash.
To obtain a $convective$ model in a nuclear flash, 
we first in the radiative models determine the location of the base of a convection zone
as the inner most layer at which the condition for unstable helium burning
$$
{\partial \epsilon_{3\alpha}\over\partial\ln T}\ge2\epsilon_{CNO}^{max}
\left[1-{1\over 4}\left({\partial\ln\kappa\over\partial\ln T}\right)_\rho\right]
\eqno (10)
$$
is satisfied (Cumming \& Bildsten 2000), where $\epsilon_{3\alpha}$
denotes the $3\alpha$ energy generation rate.
We treat the mass extent $z_c-z_e$ of the convection zone above the base as a parameter, where
$z_c$ and $z_e$ are the mass depths at the base and at the top of the convection zone,
respectively.
The temperature gradient in the convection zone is given by equation (5), and
the composition is determined by  
$\bar X_i=\int_{z_e}^{z_c} dz X_i(z)/(z_c-z_e)$, where $X_i$ in the integrand is
the composition in the corresponding radiative model.
When we integrate the basic equations from the surface to the bottom of the envelope
so that $L_r=0$ at $z=z_b$,
there appears 
a strong density discontinuity due to the discontinuous change of compositions $X_j$
at the boundaries between the convection zone and the radiative regions, at which
we assume the continuity of the pressure and the temperature.
The envelope models thus constructed to have a convective zone are 
simply called convective models in this paper.

\section{Oscillation Calculation}

For nonadiabatic and nonradial oscillations of rotating stars,
we use the method of solution given by Lee \& Saio (1987, 1993).
Throughout this paper, we apply the Cowling approximation, neglecting the Euler perturbation
of the gravitational potential, $\Psi^\prime$.
Since separation of variables is not possible for
global oscillations of rotating stars, 
the perturbations are expanded in terms of
spherical harmonic functions $Y_l^m(\theta,\phi)$ for a given $m$,
assuming axisymmetric equilibrium of the stars.
The displacement vector $\pmb{\xi}(r,\theta,\phi,t)$ is then given by
$$
\xi_r=r\sum_{l\ge|m|}^\infty S_l(r)Y_l^me^{i\sigma t}, \eqno (11)
$$
$$
\xi_\theta=r\sum_{l,l^\prime\ge|m|}^\infty \left(
H_l(r){\partial\over\partial\theta}Y_l^m+
T_{l^\prime}(r){1\over\sin\theta}{\partial\over\partial\phi}Y_{l^\prime}^m\right)
e^{i\sigma t}, \eqno (12)
$$
$$
\xi_\phi=r\sum_{l,l^\prime\ge|m|}^\infty \left(
H_l(r){1\over\sin\theta}{\partial\over\partial\phi}Y_l^m-
T_{l^\prime}(r){\partial\over\partial\theta}Y_{l^\prime}^m\right)
e^{i\sigma t}, \eqno (13)
$$
and the Euler perturbation of the pressure, $p^\prime (r,\theta,\phi,t)$,
for example, is given by
$$
p^\prime=\sum_{l\ge|m|}^\infty p_l^\prime(r)Y_l^m e^{i\sigma t}, \eqno (14)
$$
where $\sigma$ is the oscillation frequency in an inertial frame,
and $l=|m|+2(j-1)$ and $l^\prime=l+1$ for even modes, and $l=|m|+2(j-1)+1$ and $l^\prime=l-1$
for odd modes where $j=1,~2,~3,~\cdots$.
The expansions of pertubations are substituted into a set of linearized basic equations to give
a set of simultaneous linear ordinary differential equations for the expansion coefficients
$S_l(r)$, $H_l(r)$, $iT_{l^\prime}(r)$, $p_l^\prime(r)$, and so on.
In a convective region, 
the total energy flux $\pmb{F}$ is given as the sum of the radiative flux $\pmb{F}_{rad}$ 
and the convective flux $\pmb{F}_{conv}$ so that $\pmb{F}=\pmb{F}_{rad}+\pmb{F}_{conv}$.
Since there exists no reliable theory for the interaction between pulsations
and convective motion, for nonadiabatic calculations
we assume for simplicity $\delta (\nabla\cdot\pmb{F}_{conv})=0$ and $L_{rad}=L_r$
in the convective zone, where $L_r=4\pi r^2 F_r$ and $L_{rad}=4\pi r^2 F_{rad,r}$.

The inner mechanical boundary condtion we employ is given by $p^\prime=0$
(McDermott \& Taam 1987), and
the inner thermal boundary condition is $|\delta s/c_p|<<1$
(see, Lee \& Baraffe 1995), where $s$ is the specific entropy, and $c_p$ is 
the specific heat at constant pressure.
The outer boundary conditions are $\delta p=0$ and $\delta(4\pi r^2 \sigma_{SB}T^4)=\delta L_r$.
The normalization condition $\max |iT_{l^\prime}|=1$
is imposed at the surface of the envelope.
Because of the density discontinuity at the interfaces between the radiative regions 
and the convective zone, we need to impose 
jump conditions for the eigenfunctions, which are for adiabatic oscillations
$$
[\xi_r]^+_-=0, \quad [\delta p]^+_-=0, \eqno (15)
$$
where $[f(x)]^+_-=\lim_{\epsilon\rightarrow +0}(f(x+\epsilon)-f(x-\epsilon))$.
For nonadiabatic calculations, we need additional jump conditions, for which 
we employ for simplicity 
$$
[\delta L_{rad}]^+_-=0, \quad [\delta s]^+_-=0. \eqno (16)
$$

Truncation of the series expansions of perturbations
is inevitable to get a set of linear ordinary differential
equations of a finite dimesnion for numerical calculation.
We find that including the first 6 to 8 expansion coefficients is
sufficient to obtain reasonbale convergence of the eigenfrequencies
and eigenfunctions of the modes considered in this paper.

\section{Numerical Results}

\subsection{Envelope Models}

For $M=1.4M_\odot$, $R=10^6$cm, and $z_b/M_\odot=1\times 10^{-10}$ and 
for the surface chemical composition $X=0.7$, $Z=0.02$, and $Z_{CNO}=Z/2$,
we calculated envelope models for $\dot M=0.1 \dot M_{Edd}$ and 
$\dot M=0.02 \dot M_{Edd}$, where $\dot M_{Edd}$ denotes the Eddignton accretion rate
defined as $\dot M_{Edd}\equiv 4\pi c R/\kappa_e=1.88\times10^{18}(1+X)^{-1}(R/10^6)$g s$^{-1}$ and 
$\kappa_e=0.2(1+X)$ is the electron scattering opacity.
For the mass extent of the convection zone, we assume $z_e/z_c=0.01$.
In Table 1, we tabulate several physical quantities characterizing the envelopes,
such as $Ls/L_{Edd}$, the radial thickness $\Delta r$ of the envelope, the temperature $T_b$
at the bottom of the envelope, and  $y_c\equiv z_c/(4\pi R^2)$ for convective models, where
$L_{Edd}\equiv4\pi c G M/\kappa_e$ is the Eddington luminosity.
The radial extent $\Delta r$ and the base temperature $T_b$
of the convective envelope models are larger than those of the 
radiative models.
The base $z_c$ of the convective zone resides in the mixed hydrogen-helium
burning region for the model of $\dot M=0.1\dot M_{Edd}$, while 
the base is in the hydrogen-exhausted region for the model of $\dot M=0.02\dot M_{Edd}$.
In Figure 1, we plot the temperature $T$ versus the column depth
$y\equiv z/(4\pi R^2)$ for $\dot M=0.1\dot M_{Edd}$ in panel (a)
and for $\dot M=0.02\dot M_{Edd}$ in panel (b), 
where the solid lines and the dotted lines are for
the radiative and the convective models, respectively.
The temperature of the convective models is much higher than that of the
radiative models.
Both for the radiative and the convective models, the layers below the nuclear burning regions
are nearly isothermal.
To understand the modal properties of oscillation modes, it is useful to plot 
the Brunt-V\"ais\"al\"a frequency $N=\sqrt{-gA}$ with $g=GM_r/r^2$ and
the Lamb frequency $L_l=\sqrt{l(l+1)}c_s/r$ with $c_s$ being the sound velocity,
where the Schwartzschild discriminant $A$ is defined as
$$
A\equiv{d\ln \rho\over d r}-{1\over\Gamma_1}{d\ln p\over d r}=
-{\rho g\over p}\left[{\chi_T\over\chi_\rho}(\nabla_{ad}-\nabla)+\nabla_\mu\right],
\eqno (17)
$$
where
$$
\Gamma_1=\left({\partial\ln p\over\partial\ln\rho}\right)_{ad}, \quad
\chi_T=\left({\partial\ln p\over\partial\ln T}\right)_\rho, \quad
\chi_\rho=\left({\partial\ln p\over\partial\ln \rho}\right)_T, \quad
\nabla={d\ln T\over d\ln p}, \quad
\nabla_\mu={d\ln\mu\over d\ln p}, \eqno (18)
$$
and $\mu$ is the mean molecular weight.
In Figure 2, $N^2$ and $L_l^2$ with $l=1$ are plotted versus $y$ for the models
of $\dot M=0.1\dot M_{Edd}$ in panel (a) and for the models of $\dot M=0.02\dot M_{Edd}$
in panel (b), where the solid lines and the dotted lines indicate respectively the
radiative and the convective models, 
and $N^2$ and $L_l^2$ are normalized by $GM/R^3$.
Note that we have $N^2<0$ in convective regions.
Figure 3 gives the mean molecular weight $\mu$ as a function of $y$,
where the solid line and the dashed line are for
the radiative and the convective models of $\dot M=0.1\dot M_{Edd}$,
and the dotted line and the dash-dotted line are for the 
radiative and the convective models of $\dot M=0.02\dot M_{Edd}$.
Figure 3 shows
that the spiky or needle-like features found in
the Brunt-V\"ais\"al\"a frequency is produced by
extremely steep increase of the mean molecular weight $\mu$ with increasing $y$ 
in the active nuclear burning regions.

\begin{deluxetable}{cccccc}
\footnotesize
\tablecaption{Physical parameters of the envelope models}
\tablewidth{0pt}
\tablehead{\colhead{$\dot M/\dot M_{Edd}$} & \colhead{$L_s/L_{Edd}$} &
\colhead{$\Delta r~$(cm)} & \colhead{$T_b~$(K)} & 
\colhead{$y_c~$(g/cm$^2$)} & \colhead{\rm comment}}
\startdata
  0.1& 2.70E$-$3 & 3.17E3 & 4.02E8 & $\cdots$ &{\rm radiative} \\
     & 1.75E$-$2 & 4.55E3 & 9.70E8 & 1.50E8 & {\rm convective} \\
0.02 & 5.26E$-$4 & 2.61E3 & 2.01E8 & $\cdots$ & {\rm radiative} \\
     & 4.34E$-$1 & 6.82E3 & 1.45E9 & 1.65E8 & {\rm convective} \\
\enddata
\label{}
\end{deluxetable}

\begin{deluxetable}{ccccccc}
\footnotesize
\tablecaption{Oscillation frequency $\nu_{l^\prime}^a=\omega_{rad}/2\pi$ of the
fundamental $r$-modes of the radiative envelopes}
\tablewidth{0pt}
\tablehead{\colhead{$\dot M/\dot M_{Edd}$} & \colhead{$m$} &
\colhead{$\Omega/2\pi$$^a$} & \colhead{$\nu_{l^\prime=|m|}$} & 
\colhead{$\nu_{l^\prime=|m|+1}$} & \colhead{$\nu_{l^\prime=|m|+2}$} & 
\colhead{$\nu_{l^\prime=|m|+3}$}}
\startdata
  0.1& 1 & 300 & 128.4 & 11.51 & 7.076 & 5.160 \\
     &   & 600 & 188.1 & 11.52 & 7.002 & 5.062 \\
     & 2 & 300 & 113.6 & 21.79 & 13.61 & 9.995 \\
     &   & 600 & 172.6 & 22.45 & 13.75 & 9.974 \\
 0.02& 1 & 300 & 112.4 & 8.543 & 5.219 & 3.784 \\
     &   & 600 & 163.8 & 8.558 & 5.193 & 3.800 \\
     & 2 & 300 & 101.2 & 16.42 & 10.15 & 7.404 \\
     &   & 600 & 152.1 & 16.78 & 10.24 & 7.469 \\
\enddata
\label{}
\tablenotetext{a}{Frequencies are given in Hz}
\end{deluxetable}

\subsection{Eigenfrequencies}

It is now well know that the corotating-frame oscillation frequency  
$\omega\equiv\sigma+m\Omega$ of the $r$-modes
in non-isentropic stars is in the limit of $\Omega\rightarrow 0$ 
asymptotic to the frequency given by
$$
\omega=2m\Omega/[l^\prime(l^\prime+1)], \eqno (19)
$$ 
and that for a given combination of $(m,l^\prime,\Omega)$ there exist
the fundamental $r$-mode
with no radial nodes of the eigenfunction $iT_{l^\prime}$ and its overtones
(see, e.g., Yoshida \& Lee 2001).
Note that the inertial-frame oscillation frequency $\sigma$ of the $r$-modes
is negative for positive $m$, indicating the modes are prograde observed in an inertial frame.
In this paper,
we calculate the fundamental $r$-modes of $l^\prime=|m|, ~|m|+1, ~|m|+2, ~|m|+3$ for
$m=1$ and $m=2$ at several values of $\Omega$ for the envelope models.
In Figure 4, the oscillation frequencies $\omega/2\pi$ for the radiative models
are plotted versus $\Omega/2\pi$ for 
$\dot M=0.1\dot M_{Edd}$ in panel (a) and 
for $\dot M=0.02\dot M_{Edd}$ in panel (b), where
the solid lines with open symbols and those with filled symbols are 
for the modes of $m=1$ and $m=2$, respectively, and the circles, squares, triangles, and
diamonds indicate the $r$-modes of $l^\prime=|m|$, $|m|+1$, $|m|+2$, and $|m|+3$,
respectively.
As $\Omega$ is increased, 
the frequency $\omega$ of the $r$-modes of $l^\prime>|m|$
deviates appreciably from
the asymptotic frequency (19), and becomes almost insensitive to $\Omega$, gathering
in a frequency range of $\omega/2\pi\ltsim 10$Hz, at rapid rotation rates.
For convenience, we have tabulated in Table 2 the oscillation frequency of the
fundamental $r$-modes of $m=1$ and 2 for $\Omega/2\pi=300$ Hz and 600 Hz
for the radiaitve envelope models.
Table 2 clearly shows that the $r$-mode frequency of the radiative models 
is also dependent on the mass accretion rates and 
is practically independent of the spin frequency at rapid rotation rates 
for $l^\prime>|m|$.
As shown in the next subsection, 
the amplitudes of the $r$-modes at rapid rotation rates are strongly confined in the equatorial
regions of the star, in which
the Coriolis parameter $2\Omega\sin\hat\theta$ with
$\hat\theta$ being the latitude has small values.
The modal properties found for the $r$-modes of a thin spherical shell at rapid rotation rates
are quite similar to those of the low frequency equatorial waves in the ocean of the earth 
(e.g., Pedlosky 1987).

Since in the propagation zone of the $r$-modes 
the pressure scale height $H_p\approx y/\rho$ of the convective models is
larger than that of the radiative models because of higher envelope
temperature of the former, the oscillation frequency $\omega$ of the $r$-modes in the
convective models is increased compared to that in the radiative models
(see the dispersion relation given in Discussion).
This increase in $\omega$ for the convective models
in turn leads to a $decrease$ in the inertial-frame oscillation frequency
$|\sigma|$, compared to that for the radiative models.
In Figure 5, the relative difference $f$ of the oscillation frequency 
of the fundamental $r$-modes between the radiative and the convective models
is plotted versus $\sigma_{rad}/2\pi$
for $\dot M/\dot M_{Edd}=0.1$ in panel (a) and
for $\dot M/\dot M_{Edd}=0.02$ in panel (b), where
$$
f=-(\sigma_{conv}-\sigma_{rad})/ \sigma_{rad}, \eqno (20)
$$
and $\sigma_{conv}$ and $\sigma_{rad}$ are the oscillation frequencies for the
convective and the radiative envelope models, respectively, and
the solid lines with open symbols and those with filled symbols are 
for the $r$-modes of $m=1$ and
$m=2$, respectively, and the circles, squares, triangles, and
diamonds indicate the $r$-modes of $l^\prime=|m|$, $|m|+1$, $|m|+2$, and $|m|+3$,
respectively.
Note that the relative frequency difference $f$ 
for the $r$-modes of $(m,l^\prime)=(1,1)$ is larger than 0.1,
and is outside the range employed for the figure.
As $-\sigma_{rad}$ is increased, $f$ for the fundamental $r$-mode
of given $(m,l^\prime)$ decreases gradually after passing a maximum.
We find that
the relative differences $f$ at $|\sigma_{rad}|/2\pi\sim$ 300Hz for $m=1$ and at
$|\sigma_{rad}|/2\pi\sim$ 600Hz for $m=2$, for example, have similar values, and that
$f$ for given $m$ and $\Omega$ becomes smaller for larger values of $l^\prime$.
If we compare the relative differences $f$ calculated here with the frequency drifts 
observed in burst oscillations in X-ray bursters,
we find that the fundamental $r$-modes
of $l^\prime>|m|+1$ produce relative differences $f$ consistent with observed frequency drifts
at $|\sigma|/2\pi\sim300$Hz for $m=1$ or at $|\sigma|/2\pi\sim600$Hz for $m=2$,
although the differences $f$ for the $r$-modes of $l^\prime=|m|$ are too large to be
reconciled with the observations for the models used in this paper.
If the star is rotating as rapidly as $\Omega/2\pi\sim600$Hz, $f$ for the
$m=1$ $r$-modes also becomes consistent with observed values, which suggests 
the possibility that the burst QPO near 600Hz could be produced by the 
$r$-modes of $m=1$.
We also note that to obtain values of $f$ as small as $\sim$0.001 we need to consider
the $r$-modes that have the index $l^\prime$ much larger than $|m|$.

Although the $r$-modes in the radiative models are all pulsationally stable
for the mass accretion rates considered in this paper (see Strohmayer \& Lee 1996),
the fundamental $r$ modes in the convective models are driven unstable by
strong nuclear burning in the convective zone.
In Figure 6, the growth timescale  
$\tau_{growth}\equiv-1/\sigma_I$ of the $r$-modes in the convective models
is plotted as a function of $\Omega/2\pi$
for $\dot M=0.1\dot M_{Edd}$ in panel (a) and 
for $\dot M=0.02\dot M_{Edd}$ in panel (b), where
$\sigma_I$ is the imaginary part of the eigenfrequency $\sigma$, and 
the solid lines with open symbols and those with filled symbols are for the modes of $m=1$
and $m=2$, respectively, and the circles, squares, and triangles
indicate the $r$-modes of $l^\prime=|m|$, $|m|+1$, and $|m|+2$, 
respectively.
$\tau_{growth}$'s for $\dot M=0.02\dot M_{Edd}$ are by about three
order of magnitude shorter than those for $\dot M=0.1\dot M_{Edd}$,
because the envelope temperature of the former is much higher
than that of the latter.
Note that $\tau_{growth}$ becomes almost insensitive to $\Omega$ for rapid rotation rates.

\subsection{Eigenfunctions}

The expansion coefficients $iT_{l^\prime}$ and $H_l$ of the fundamental $r$-modes
of $(m,l^\prime)=(2,3)$ at $\bar\Omega=\Omega/\sqrt{GM/R^3}=0.1$
are shown versus the column depth $y$ 
for the models with $\dot M=0.1\dot M_{Edd}$ in Figure 7, and those of $(m,l^\prime)=(1,1)$
for the models with $\dot M=0.02\dot M_{Edd}$ in Figure 8, where 
the panels (a) and (b) are for the radiative models
and the panels (c) and (d) for the convective models, and the amplitude
normalization is given by $\max |iT_{l^\prime}|=1$ at the surface.
As shown by Figure 7,
the component $iT_{l^\prime=|m|+1}$ of the $r$-mode at $\bar\Omega=0.1$
is no longer dominant over the other components of $iT_{l^\prime}$ and $H_l$.
This is also the case for the component $iT_{l^\prime=|m|}$ of
the $r$-mode of $(m,l^\prime)=(1,1)$ shown in Figure 8.
The modal properties of the modes at $\bar\Omega=0.1$ are thus quite different from
the properties found for the $r$-modes in the limit of $\bar\Omega\rightarrow 0$, 
for which the toroidal component $iT_{l^\prime=|m|+1}$ 
(or $iT_{l^\prime=|m|}$)
is dominant over the other components $iT_{l^\prime}$, $H_l$, and $S_l$.
This becomes clear if we show the $\theta$ dependence of the eigenfunctions
$\xi_\theta$ and $\xi_\phi$, for example.
To do so, we introduce the functions $X_\theta(r,\theta,\phi)$ and 
$X_\phi(r,\theta,\phi)$ defined as
$$
X_\theta(r,\theta,\phi) =\Re\left[ \xi_\theta(r,\theta,\phi,t)e^{-i\sigma t}/r\right],
\eqno (21)
$$
$$
X_\phi(r,\theta,\phi)=-\Re\left[i\xi_\phi(r,\theta,\phi,t)e^{-i\sigma t}/r\right],
\eqno (22)
$$
and, for $r=R$ and $\phi=0$, we plot the functions $X_\theta$ (solid lines)
and $X_\phi$ (dotted lines) versus $\cos\theta$
for the fundamental $r$-modes of $(m,l^\prime)=(1,1)$, (1,2), and (1,3) in Figure 9
and for those of $(m,l^\prime)=(2,2)$, (2,3), and (2,4) in Figure 10, where
the radiative model of $\dot M=0.02\dot M_{Edd}$ has been used for the
mode calculation, and the amplitude normalization is given by $\max |iT_{l^\prime}|=1$  
at the surface.
Here, in each figure the panels on the left-hand-side are for $\bar\Omega=0.01$,
and those on the right-hand-side are for $\bar\Omega=0.1$.
Although the $r$-modes have large amplitudes at the poles at $\bar\Omega=0.01$,
the amplitudes of the modes at $\bar\Omega=0.1$ become well confined
into the equatorial regions of the star, and have effectively no amplitudes
at the poles.
Note that the parity of the function $X_\theta\pmb{e}_\theta$ becomes the same
as that of $X_\phi$ with respect to $\cos\theta$.

It may be instructive to give a plot of $\delta L_r$ for the $r$-modes
considered in this paper,
although nonadiabatic calculation is not necessarily quite reliable
for the convective models because of our lack of knowledge concerning the interaction
between pulsations and convection, and because of somewhat adhoc way of construction of
the models.
For the convective model of $\dot M/\dot M_{Edd}=0.1$,
Figure 11 shows the expansion coefficients $\Re(\delta L_{r,l}/L_s)$
as a function of $y$ for the fundamental $r$-mode of $(m,l^\prime)=(2,3)$ in panel (a)
and for that of $(m,l^\prime)=(2,5)$ in panel (b)
at $\bar\Omega=0.1$, where 
the solid, dotted, dashed, and dash-dotted lines are for the expansion coefficients
with $l=|m|$, $|m|+2$, $|m|+4$, and $|m|+6$, respectively, and 
the amplitude normalization is given by $\max |iT_{l^\prime}|=1$ at
the surface.
Figure 12 illustrates $\Re(\delta L_{r}/L_s)$ versus $\cos\theta$, 
where the solid and the dotted lines respectively designate 
the $r$-modes of $(m,l^\prime)=(2,3)$ 
and $(m,l^\prime)=(2,5)$ at $\bar\Omega=0.1$, and we have assumed $r=R$ and $\phi=0$.
From Figures 11 and 12, we find that
$\Re(\delta L_{r,l}/L_s)$ at the surface has large amplitudes,
comparable to $\max |iT_{l^\prime}|$, in the equatorial regions,
and that, although $\Re(\delta L_{r}/L_s)$ of the $r$-mode of $(m,l^\prime)=(2,3)$
does not change its sign as a function of $\cos\theta$, 
$\Re(\delta L_{r}/L_s)$ of the $r$-mode of $(m,l^\prime)=(2,5)$ changes its sign
at a middle latitude, which may result in a reduction of the detectability of 
the signals produced by the mode.
Note that the $r$-modes of $(m,l^\prime)=(2,3)$ and $(m,l^\prime)=(2,5)$
are even modes, and 
the perturbation $\delta L_{r}/L_s$ is symmetric about the equator.
On the other hand, 
the odd $r$-modes of $l^\prime=|m|, ~|m|+2, ~\cdots$ have the eigenfunctions
antisymmetric about the equator.

\section{Discussion}

Erecting at the equator a cartesian coordinate system in which the
coordinates $x$, $y$, and $z$ are directed to the east, the north, and the zenith,
respectively, and applying the $\beta$-plane approximation at the origin, 
we may obtain
the dispersion relation for equatorial waves given by (Pedlosky 1987)
$$
\lambda\bar\omega^2+{\bar k\over\bar\omega}-{\bar k}^2=(2j+1)\sqrt{\lambda},
\eqno (23)
$$
where $\bar k$ is the dimensionless wavenumber in the $x$ direction, and
$\lambda$ is the separation constant between the horizontal and vertical momentum equations, 
and the velocity perturbation $v_y^\prime$, for example, have been
assumed to have the form given by $e^{i(kx+\omega t)}\Psi(y)G(z)$.
Here, the separation constant $\lambda$ is determined by the vertical-structure equation
that governs the function $G(z)$, and
the non-negative-integer $j$ comes from
the normal mode equation for $v_y^\prime$, for which
the function $\Psi(y)$ is shown to be
proportional to $e^{-\eta^2/2}H_j(\eta)$, where 
$H_j(\eta)$ is the Hermite polynomial and $\eta=\sqrt{\lambda}y/L_e$ (Pedlosky 1987).
In the dispersion relation, the dimensionless oscillation frequency $\bar\omega$ and 
the wavenumber $\bar k$ in the $x$ direction are normalized as
$$
\bar\omega={\omega/(\beta_0L_e)}, \quad
\bar k=L_e  k, \eqno (24)
$$
where $\beta_0={2\Omega/ R}$, and 
$$
L_e=\sqrt{N_0D/\beta_0} \eqno (25)
$$
is the equatorial Rossby internal-deformation radius, and 
$N_0$ is the characteristic value of the Brunt-V\"ais\"al\"a frequency,
$D$ is the depth of the atmosphere or the ocean (Pedlosky 1987).
Rewriting the dispersion relation using dimensional $\omega$ and $k$, and taking
the low frequency limit, we obtain
$$
{N_0D\over\omega}k-{N_0D\over\beta_0}k^2=(2j+1)\sqrt{\lambda}, \eqno (26)
$$
which can be solved for $\omega$ to give
$$
\omega={N_0Dk\over (2j+1)\sqrt{\lambda}+N_0Dk^2/\beta_0} 
\sim{mN_0(D/R)\over (2j+1)\sqrt{\lambda}+m^2N_0(D/R)/2\Omega},
\eqno (27)
$$
where we have assumed $k\sim m/R$ with $m$ being an integer corresponding to
the azumutal wave number.
Note that the oscillation frequency is proportional to the depth $D$ of the 
atmosphere or the ocean, and $D/R<<1$.
In the limit of $\Omega\rightarrow0$, we obtain
$$
\omega\rightarrow 2\Omega/m, \eqno (28)
$$
and we obtain at rapid rotation speeds
$$
\omega\sim {mN_0(D/R)\over (2j+1)\sqrt{\lambda}}, \eqno (29)
$$
which does not depend on $\Omega$.
If we regard $D$ and $N_0$ as the pressure-scale-height $H_p$ and the 
characteristic value of $N$ found in the $r$-mode propagation region,
the dispersion relation given above describes qualitatively well the
$\Omega$ dependence of the $r$-mode frequency $\omega$ obtained in this paper.
The quantity $N_0(D/R)$ in equation (29)
may depend on parameters such as
the surface gravity $g=GM/R^2$, the radius $R$, and the mass accretion rate, and
extensive model calculations would be required to see how the quantity $N_0(D/R)$
depends on the parameters.
Note that the equatorial waves have large amplitudes at the equatorial regions but
vanishing amplitudes at the poles as indicated by the
function $\Psi(y)\propto e^{-\eta^2/2}H_j(\eta)$, 
the propeties of which are similar to those of the $r$-modes
found in this paper.

\section{Conclusion}

In this paper, we calculated the $r$-modes propagating in mass-accreting, nuclear burning,
and geometrically thin envelopes on the surface of rotating neutron stars.
In steady state approximation, we constructed thin envelope models which
are fully radiative or have a convective region. 
The radiative and the convective models were intended to represent envelope
structures in a late phase and in an early phase of an X-ray burst, respectively.
The convective models have higher temperature and 
larger geometrical thickness than the corresponding radiative models.
We find that
the corotating-frame oscillation frequency $\omega$ of the fundamental 
$r$-modes of $(m,l^\prime)$, which
is asymptotic to 
$\omega=2m\Omega/[l^\prime(l^\prime+1)]$ when $\Omega\rightarrow 0$, 
becomes essentially insensitive to $\Omega$ and gathers 
in a frequency range of $\omega/2\pi\ltsim 10$Hz
at rapid rotation rates.
It is also found that the amplitudes of the $r$-modes at rapid rotation rates 
become well confined to the equatorial regions of the star.
Although the fundamental $r$-modes of the radiative envelope models are all
pulsationally stable for the mass accretion rates considered in this paper,
the $r$-modes of the convective models are driven unstable by nuclear burning
in the convection zone.

It is found that the inertial-frame oscillation frequency $|\sigma|$ of the $r$-modes
of given $(m,l^\prime,\Omega)$ 
for the convective models becomes lower than that for the corresponding radiative models. 
In order to estimate possible magnitudes of the relative frequency changes of the $r$-modes
during an X-ray burst, we calculated
$f=-(\sigma_{conv}-\sigma_{rad})/ \sigma_{rad}$ 
as a function of $\Omega$, where $\sigma_{rad}$ and $\sigma_{conv}$ are the oscillation
frequencies of the fundamental $r$-modes in the radiaitve and the convective envelope
models, respectively.
We find that, if we consider the burst oscillations of $|\sigma_{rad}|/2\pi\sim$ 300Hz 
or $|\sigma_{rad}|/2\pi\sim$ 600Hz, the fundamental $r$-modes of $l^\prime > |m|+1$ 
with $m=1$ or $m=2$
can produce the relative frequency changes $f$ consistent with
the observed relative frequency drifts of less than $\sim$1\%, although we need to consider
the $r$-modes of $l^\prime$ much larger than $|m|$ for values of $f$ as small as $\sim$0.001.
We also find that $f$ for the $\dot M=0.02\dot M_{Edd}$ models 
is larger than that for the $\dot M=0.1\dot M_{Edd}$ models,
which may reflect the properties of nuclear flashes dependent on the mass accretion rates
(see, e.g., Fujimoto, Hanawa, \& Miyaji 1981) and could be tested observationally.
Considering that the fundamental $r$-modes in the convective models are driven unstable by
nuclear burning in the convective zone,
it is tempting to postulate that the burst oscillations observed in X-ray bursts are produced by the
$r$-modes of low indices $(m,l^\prime)$ that are excited in the early phase of 
the nuclear flash and survive until the tail, changing the frequency slightly according to
the changes in the envelope structure.
The $r$-mode scenario is consistent with
the observations that indicate the existence of two frequencies separated by
a few Hz, since the fundamental $r$-modes
of different $l^\prime$'s for a given $m$ could be excited simulataneously in a burst to 
have similar oscillation frequencies for rapidly rotating neutron stars.
The temporal frequency decrease observed in the tail of bursts from 4U 1636-53 (Strohmayer 1999)
and KS 1731-260 (Muno et al 2000) may be attributable to switching of the $r$-modes
observable to us from those of ($m$,$l^\prime_1$) to 
($m$,$l^\prime_2$) where $l^\prime_1>l^\prime_2$ for frequency decrease.
The mode switching will cause discontinuous frequency changes and
could be tested by a close observation of the frequency evolution.
To make possible the close comparison between obervations and 
the $r$-mode scenario for burst oscillations, however,
it is definitely necessary to use more realistic 
envelope models obtained from time-dependent flash calculation with
a reliable treatment of the convection zone.

Recently, Chakrabarty et al (2003) reported the detection of quasi-periodic oscillatons
in X-ray bursts of the millisecond pulsar SAX J1808.4-3658, showing that
the oscillation frequencies agree quite well with the 
spin frequency of the star.
The time evolution of the burst oscillation frequencies, however, was found 
quite different from that in other X-ray bursters that show burst oscillations.
The frequency drift $\Delta\nu/\nu\sim 0.01$ is among the largest observed in any neutron star.
The drift time scale is an order of magnitude faster than in the other neutron stars, and
the oscillation frequency overshoots the spin frequency, 
reaching the maximum during the burst rise.
The burst oscillations found in the pulsar are inconsistent with both the 
contracting shell model with angular momentum conservation
and the $r$-mode scenario.
A key factor for the burst oscillaitons in the star 
may be a strong magnetic field infered from the 
existence of coherent X-ray pulsations in non-burst phases, since the strong magnetic field
has significant
effects both on the modal properties of waves propagating in the surface layers and on
the way of nuclear burning on the surface of the star.

\begin{figure}
\epsscale{.5}
\plottwo{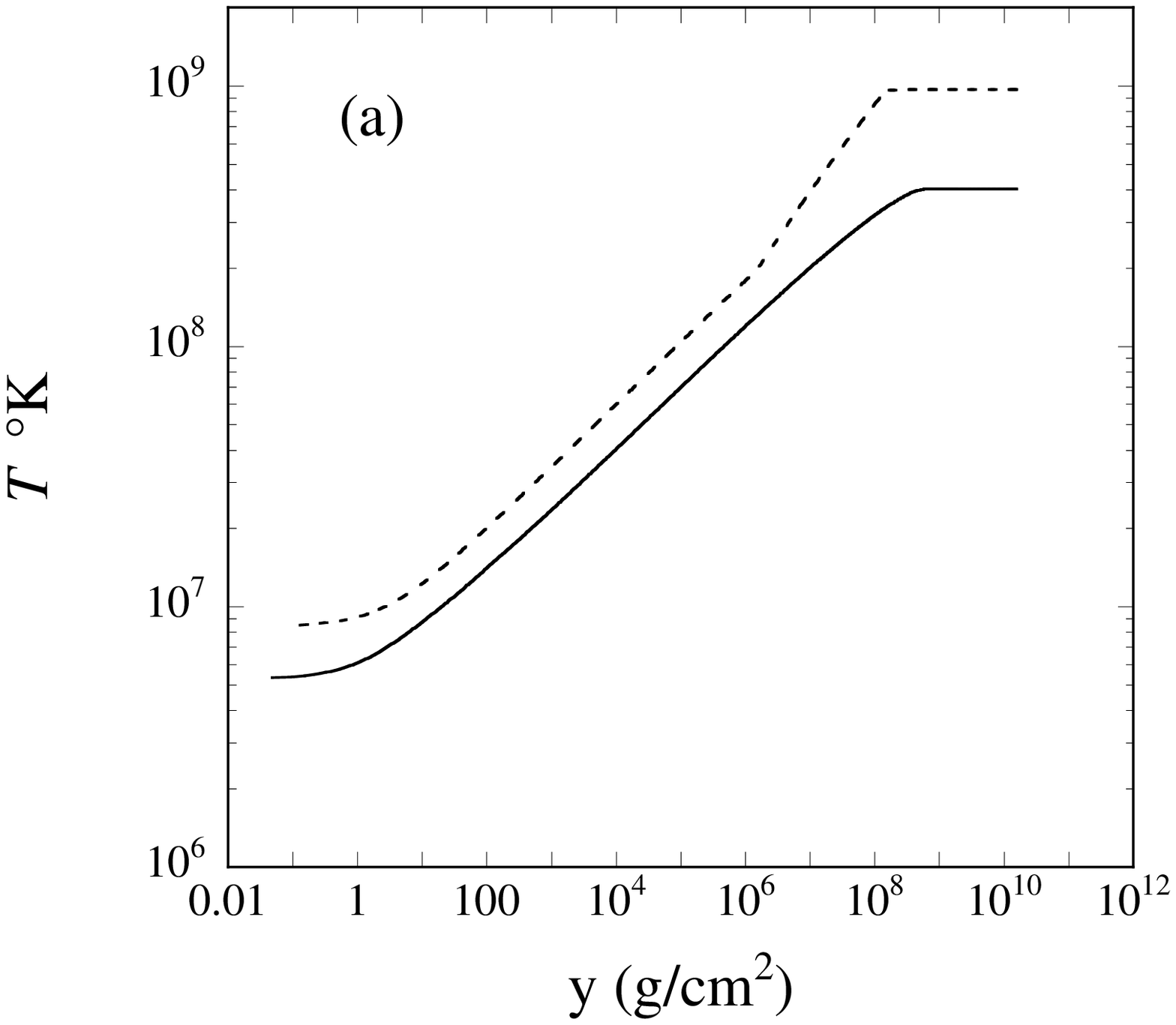}{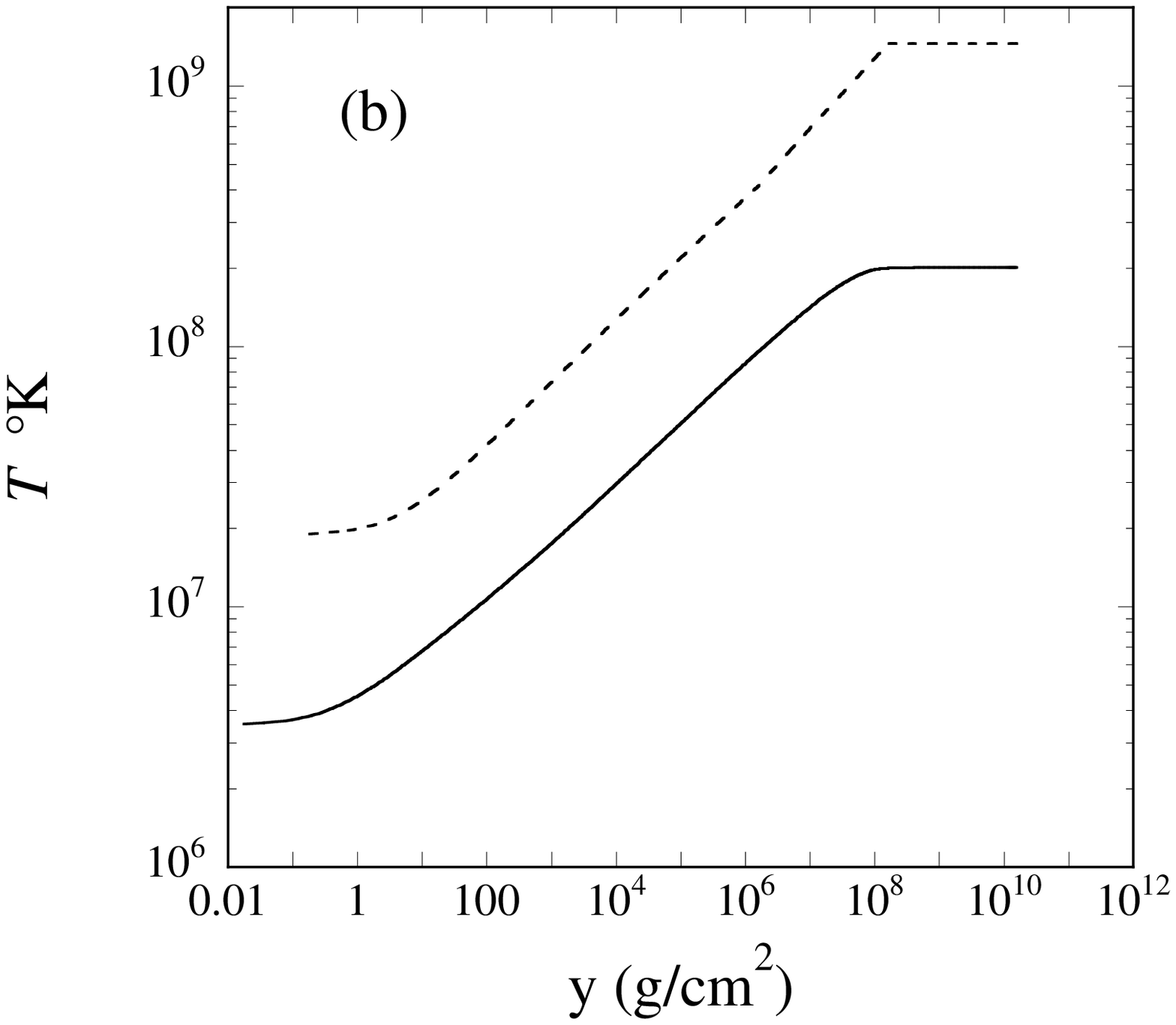}
\caption{Temperature $T$ as a function of the column depth $y$
for $\dot M=0.1\dot M_{Edd}$ in panel (a) and for $\dot M=0.02\dot M_{Edd}$ in panel (b),
where the solid lines and the dotted lines are for the radiative and the convective models,
respectively.}
\end{figure}

\begin{figure}
\epsscale{.5}
\plottwo{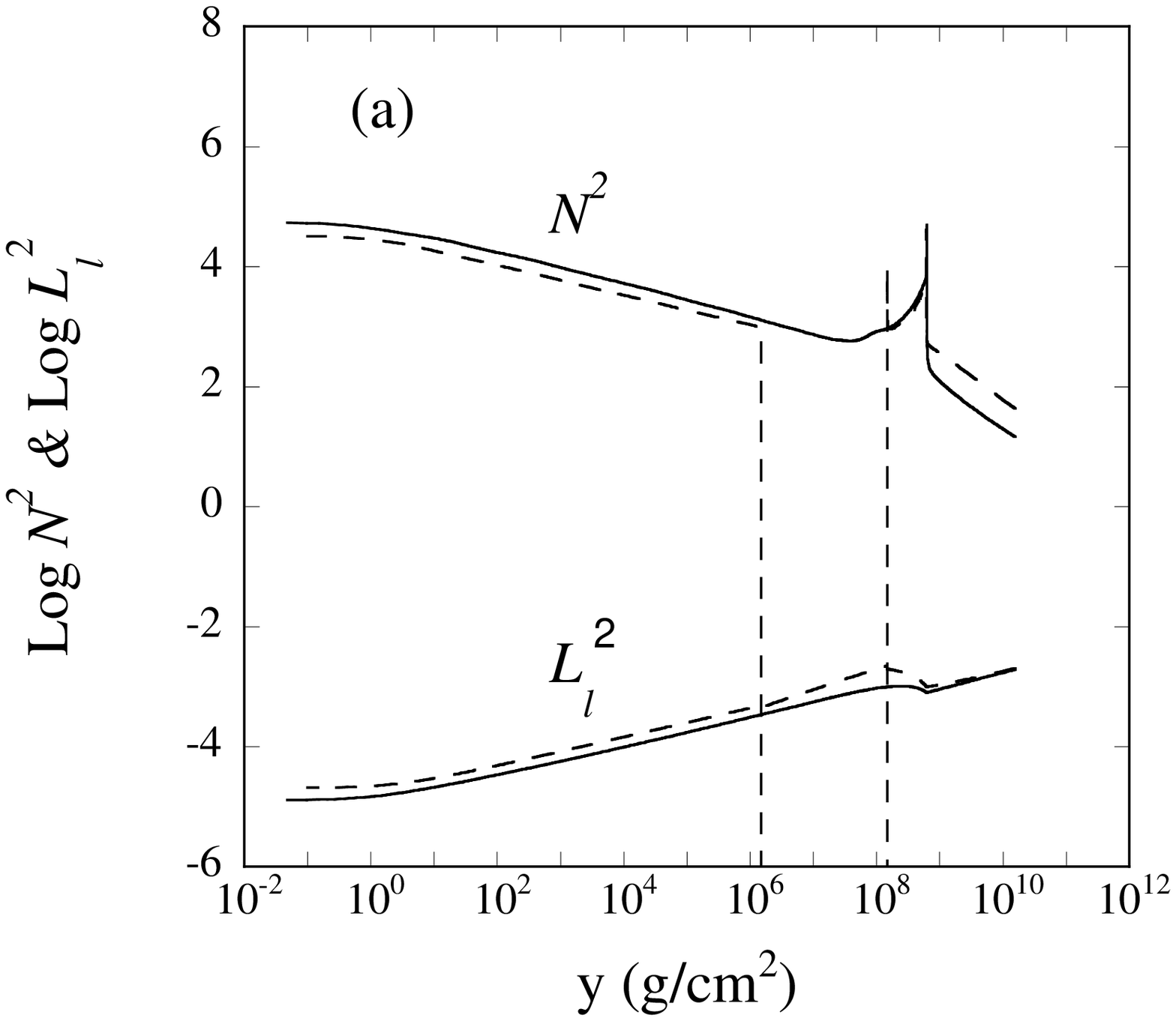}{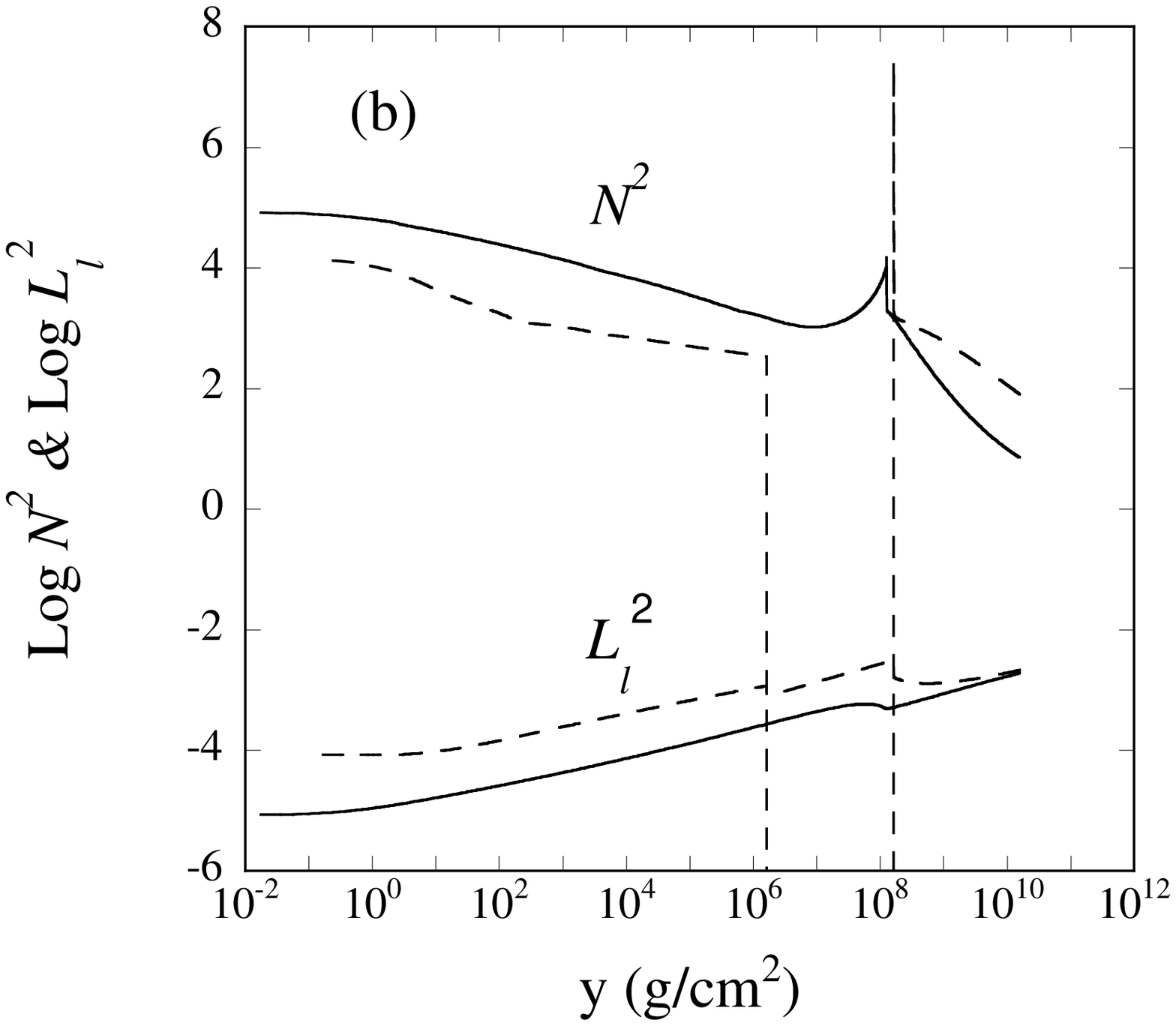}
\caption{Propagation diagram for the 
envelope models, where the square of Brunt-V\"ais\"al\"a frequency $N$ and
Lamb frequency $L_l$ with $l=1$ are displayed versus the column depth $y$
for the radiative models (solid lines) and the convective models (dashed lines).
$N^2$ and $L_l^2$ are normalized by $GM/R^3$ with $M$ and $R$ being
the mass and the radius of the neutron star. Panels (a) and (b) are for the cases
of $\dot M/\dot M_{Edd}=0.1$ and 0.02, respectively.}
\end{figure}

\begin{figure}
\epsscale{.5}
\plotone{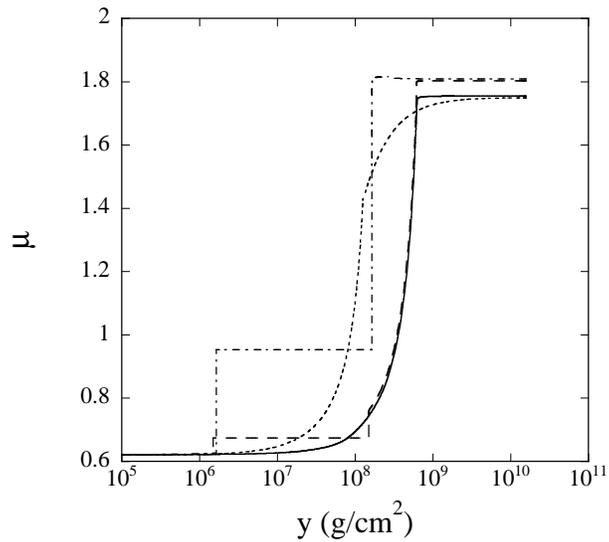}
\caption{Mean molecular weight $\mu$ as a function of the column depth $y$
for the radiative (solid line) and the convective (dashsed line) models of
$\dot M/\dot M_{Edd}=0.1$ and the radiative (dotted line) and the
convective (dash-dotted line) models of $\dot M/\dot M_{Edd}=0.02$.}
\end{figure}
\begin{figure}
\epsscale{.5}
\plottwo{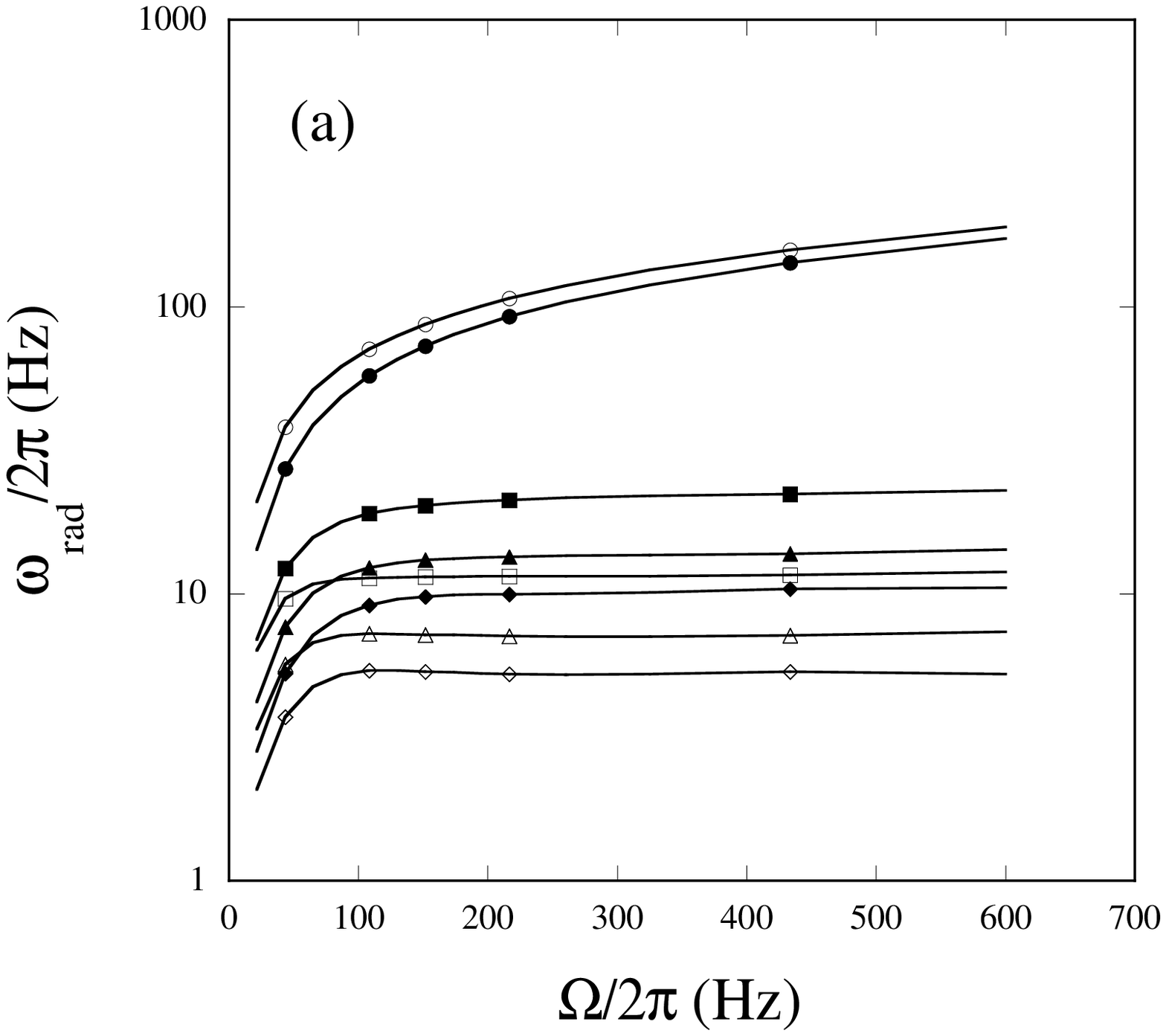}{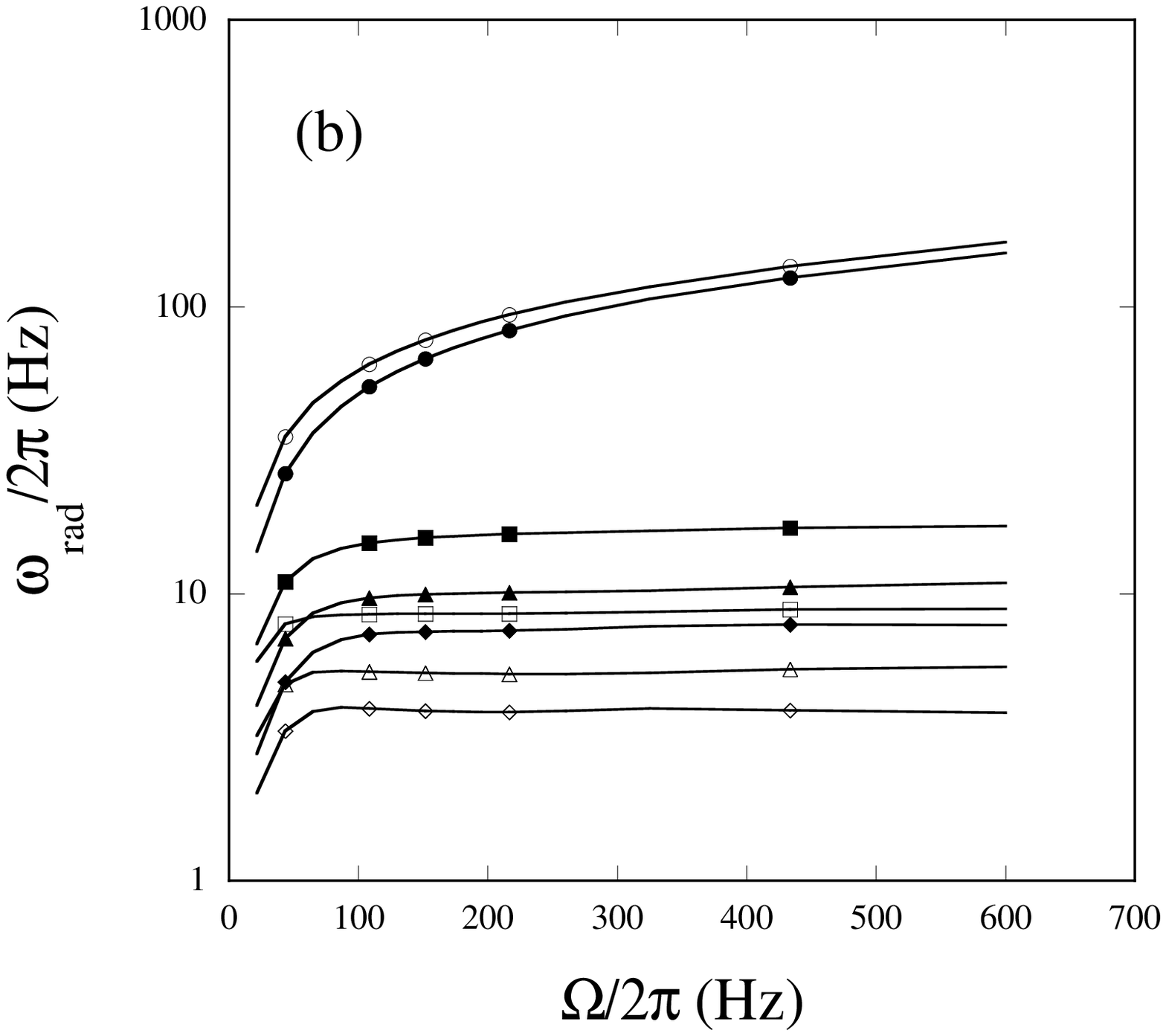}
\caption{Oscillation frequencies $\omega$ of the fundamental $r$-modes
of the radiative envelope models are given versus $\Omega/2\pi$ 
for $\dot M/\dot M_{Edd}=0.1$ in panel (a)
and for $\dot M/\dot M_{Edd}=0.02$ in panel (b), where the solid lines attached by open
symbols and those by filled symbols are for $m=1$ and $m=2$, respectively, and
the circles, squares, triangles, and diamonds indicate the $r$-modes of 
$l^\prime=|m|,~|m|+1,~|m|+2$, $|m|+3$, respectively.}
\end{figure}

\begin{figure}
\epsscale{.5}
\plottwo{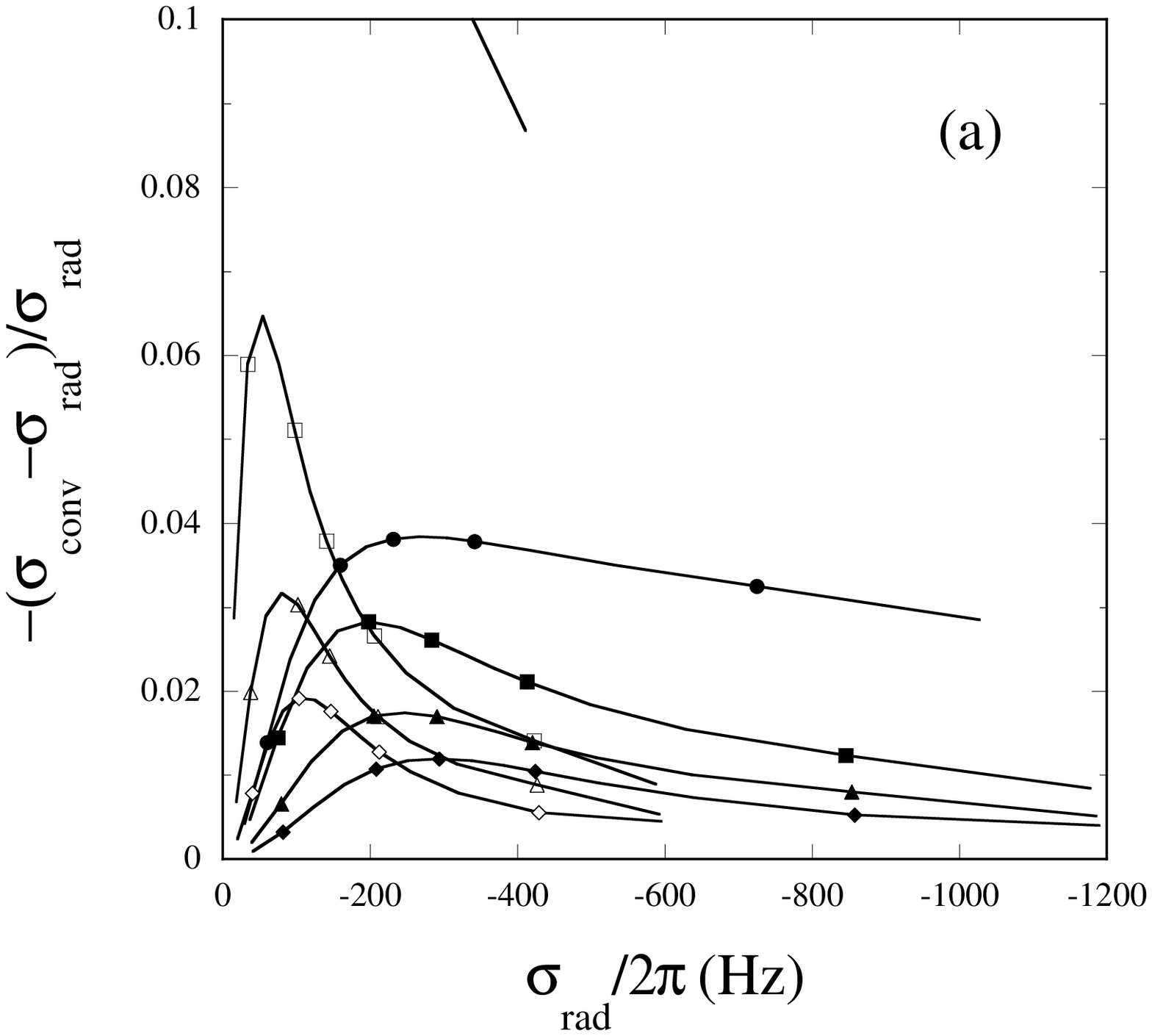}{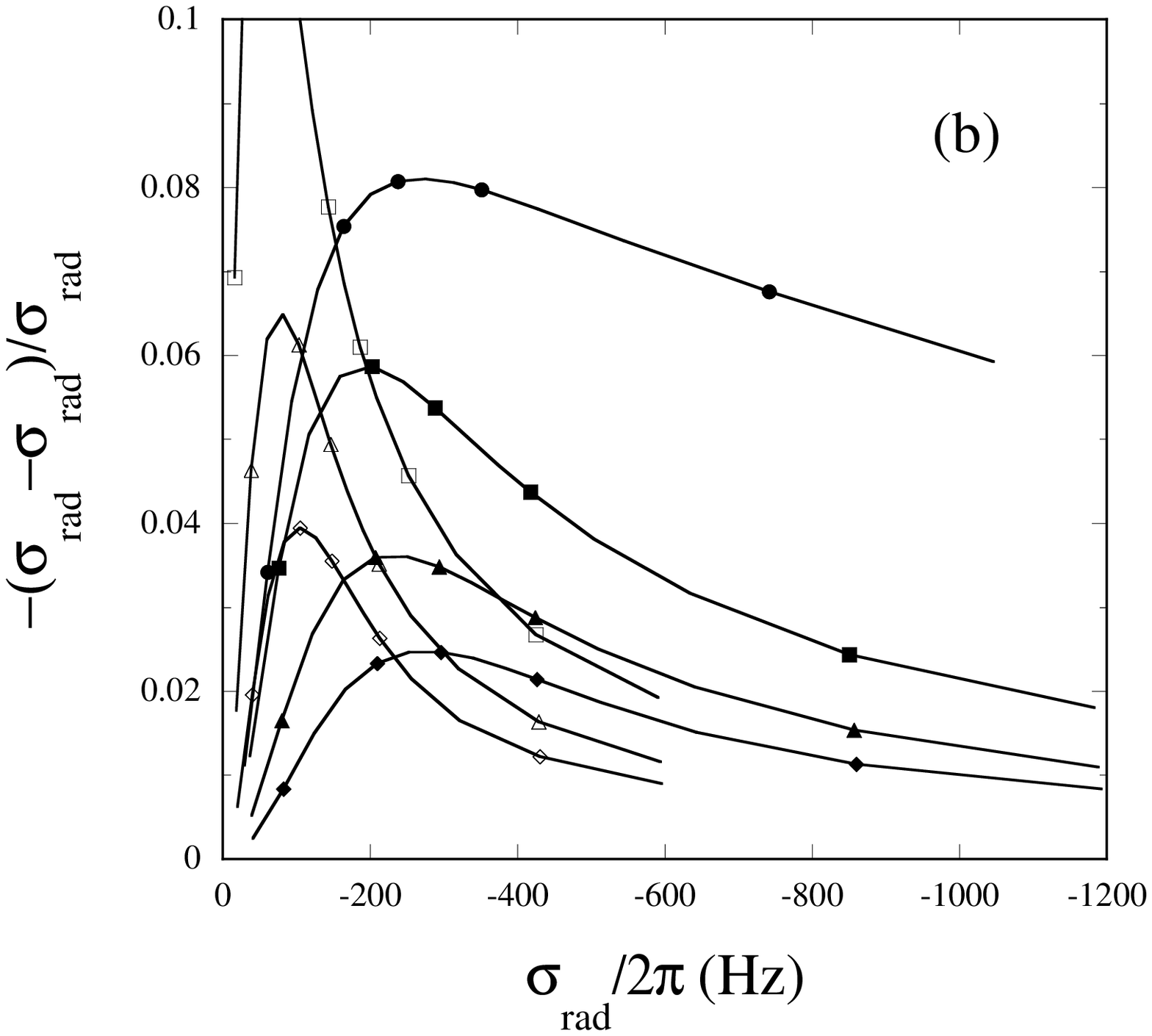}
\caption{Relative frequency differences 
$f\equiv-(\sigma_{conv}-\sigma_{rad})/\sigma_{rad}$ of the fundamental $r$-modes
between the radiative and the convective envelope models are plotted versus
$\sigma_{rad}/2\pi$
for $\dot M/\dot M_{Edd}=0.1$ in panel (a)
and for $\dot M/\dot M_{Edd}=0.02$ in panel (b), where the solid lines attached by open
symbols and those by filled symbols are for $m=1$ and $m=2$, respectively, and
the circles, squares, triangles, and diamonds indicate the $r$-modes of 
$l^\prime=|m|,~|m|+1,~|m|+2$, $|m|+3$, respectively.
Here, $\sigma_{rad}$ and $\sigma_{conv}$ are the inertial-frame oscillation frequencies
of the $r$-modes in the radiaitve and the convective models, respectively.}
\end{figure}

\begin{figure}
\epsscale{.5}
\plottwo{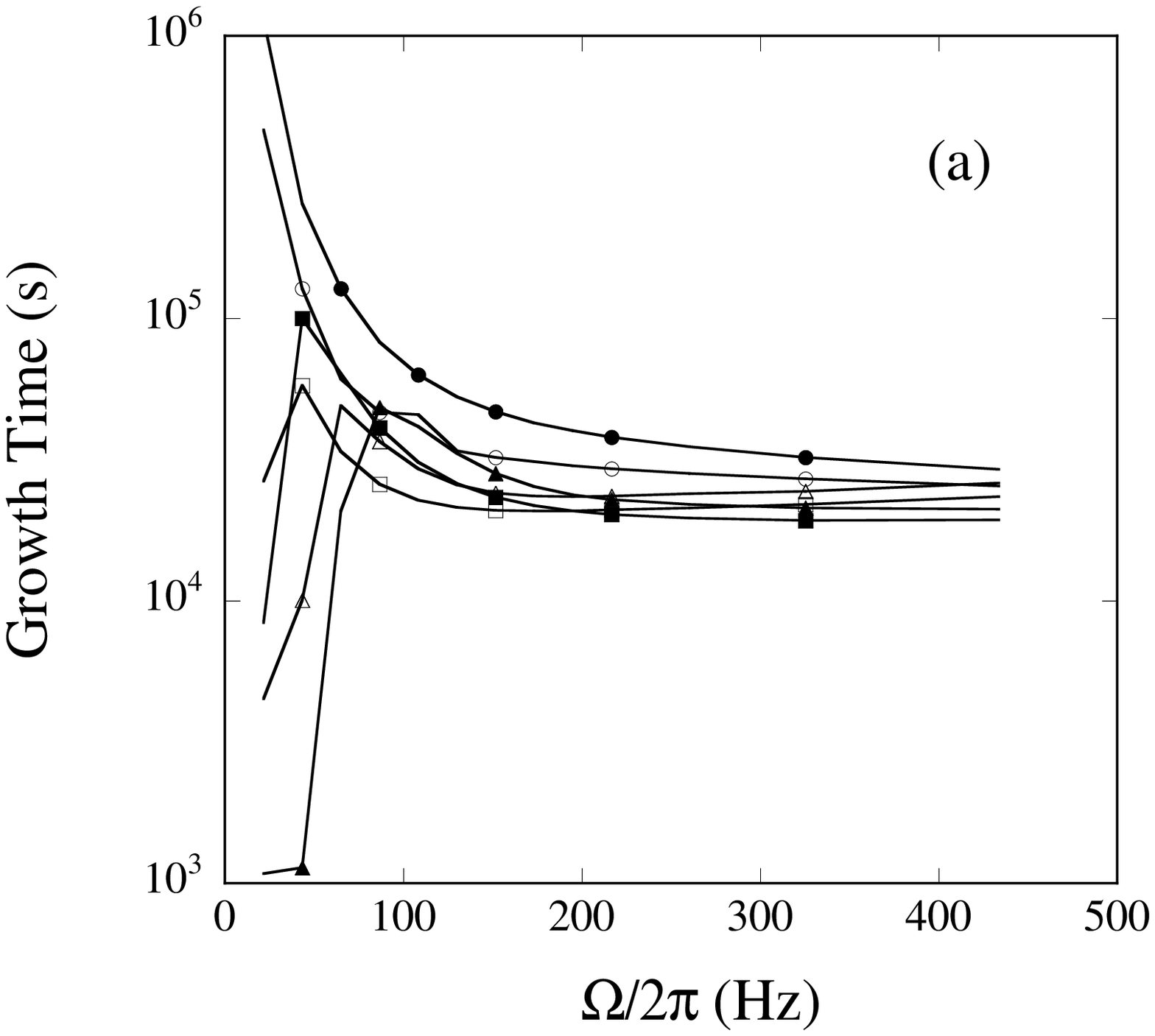}{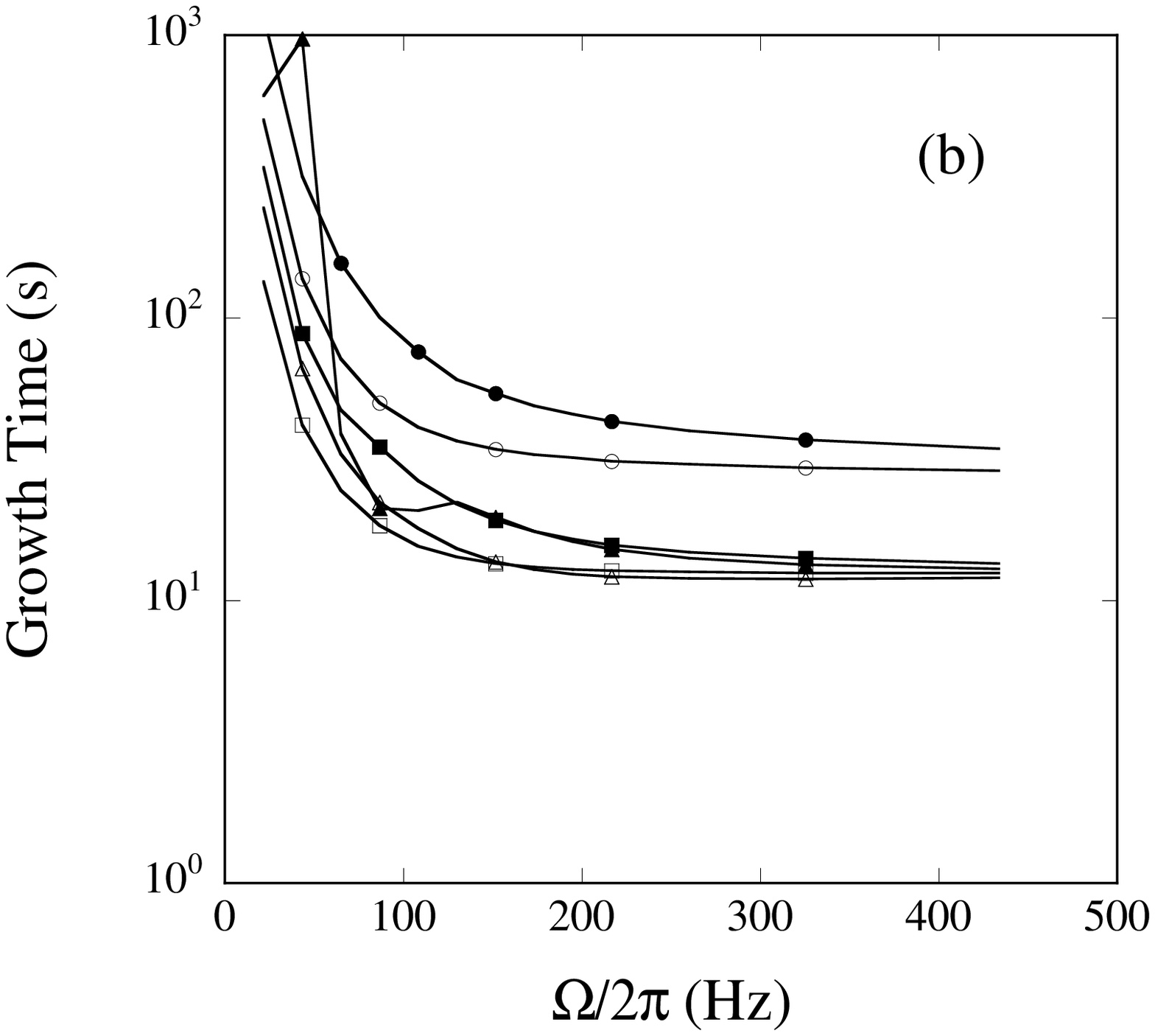}
\caption{Growth time-scales $\tau_{growth}\equiv-1/\sigma_I$
of the fundamental $r$-modes of the convective envelope models are plotted versus
$\Omega/2\pi$
for $\dot M/\dot M_{Edd}=0.1$ in panel (a)
and for $\dot M/\dot M_{Edd}=0.02$ in panel (b), where the solid lines attached by open
symbols and those by filled symbols are for $m=1$ and $m=2$, respectively, and
the circles, squares, and triangles indicate the $r$-modes of 
$l^\prime=|m|,~|m|+1,$ and $|m|+2$, respectively.
Here, $\sigma_I$ is the imaginary part of the eigenfrequency of the
$r$-modes.
}
\end{figure}

\begin{figure}
\epsscale{.5}
\plotone{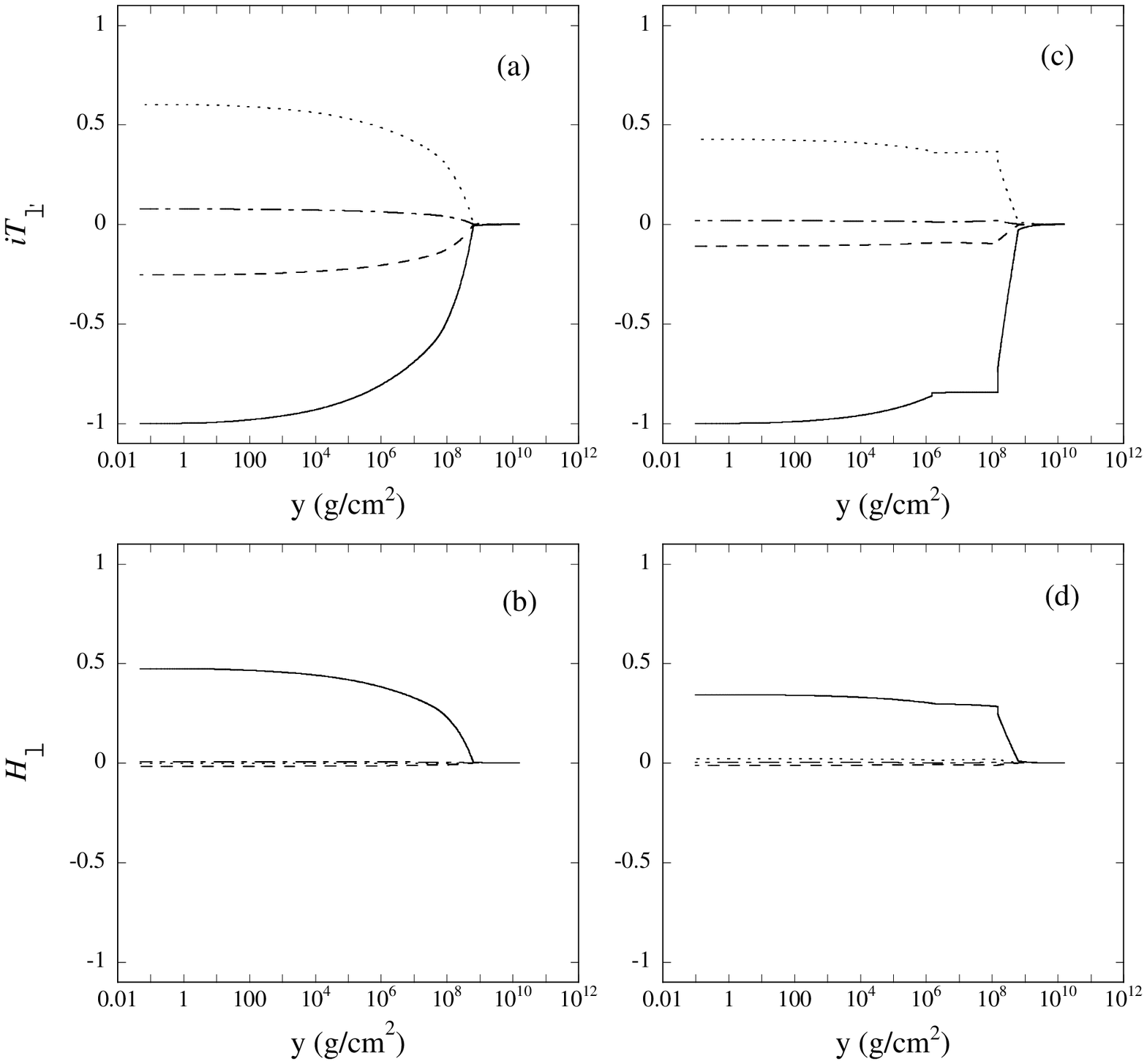}
\caption{Expansion coefficients $iT_{l^\prime}$ and $H_l$
of the fundamental $r$-mode of $(m,l^\prime)=(2,3)$ at $\bar\Omega=0.1$
are given versus the column depth $y$
for $\dot M/\dot M_{Edd}=0.1$, where the solid, dotted, dashed, and dash-dotted lines
indicate the coefficients associated with $l=l^\prime-1=|m|,~|m|+2,~|m|+4$,
and $|m|+6$, respectively, and
the amplitude normalization is given by $\max |iT_{l^\prime}|=1$ at the surface.
Here, panels (a) and (b) are for the radiative models, and panels (c) and (d)
are for the convective models.
}
\end{figure}

\begin{figure}
\epsscale{.5}
\plotone{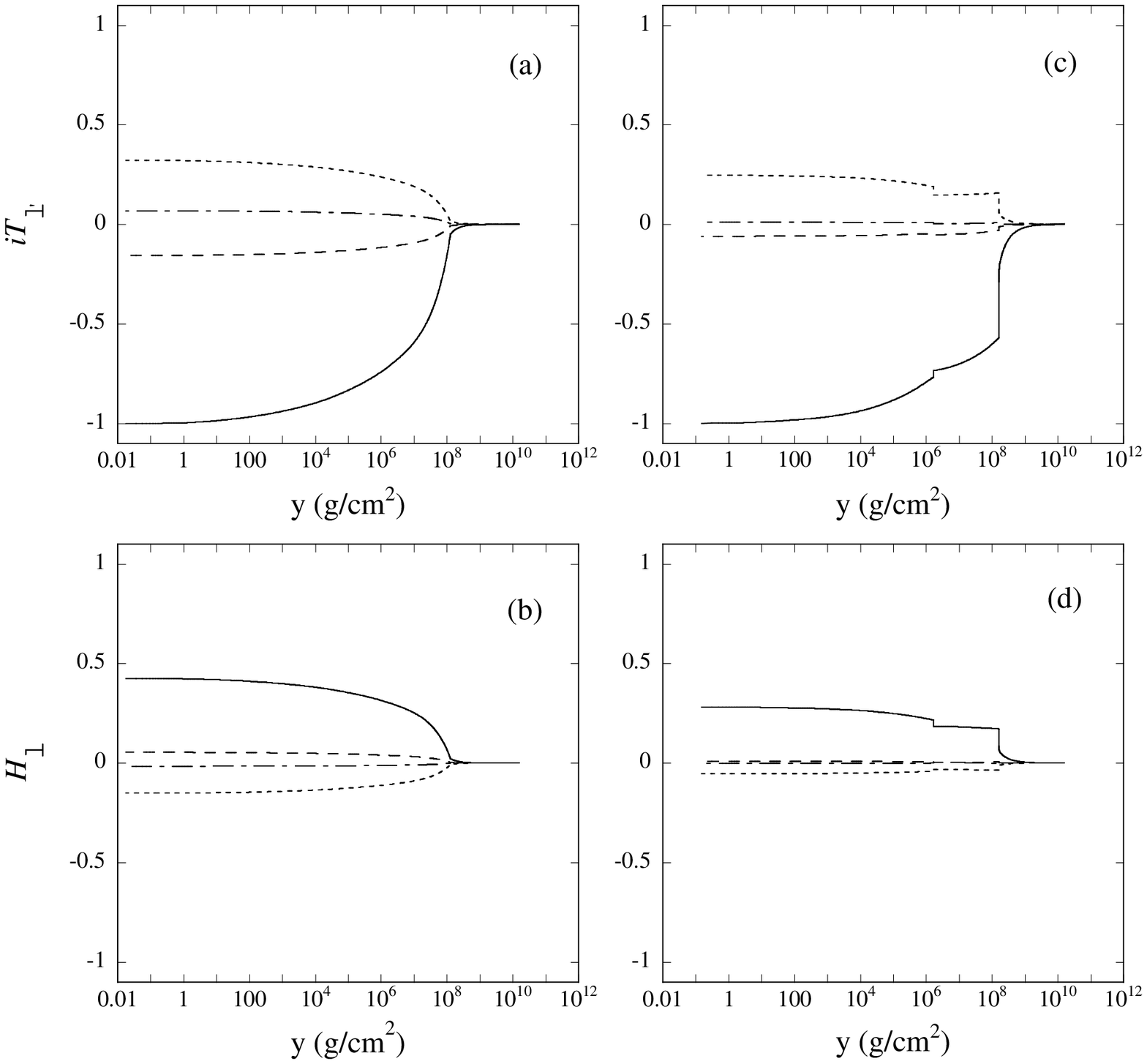}
\caption{Expansion coefficients $iT_{l^\prime}$ and $H_l$
of the fundamental $r$-mode of $(m,l^\prime)=(1,1)$ at $\bar\Omega=0.1$
are given versus the column depth $y$
for $\dot M/\dot M_{Edd}=0.02$, where the solid, dotted, dashed, and dash-dotted lines
indicate the coefficients associated with $l^\prime=l-1=|m|,~|m|+2,~|m|+4$,
and $|m|+6$, respectively, and
the amplitude normalization is given by $\max |iT_{l^\prime}|=1$ at the surface.
Here, panels (a) and (b) are for the radiative models, and panels (c) and (d)
are for the convective models.
}
\end{figure}

\begin{figure}
\epsscale{.5}
\plotone{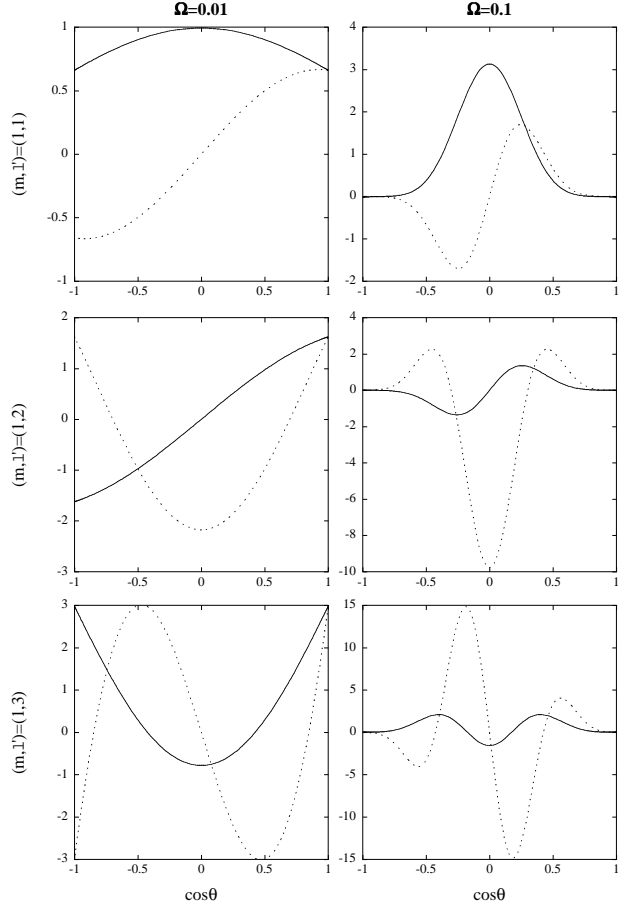}
\caption{Functions
$X_\theta(r,\theta,\phi)$ and $X_\phi(r,\theta,\phi)$ 
for $r=R$ and $\phi=0$ are plotted
versus $\cos\theta$
for the fundamental $r$-modes of $l^\prime =|m|, ~|m|+1$, and $|m|+2$ for $m=1$, where
the solid lines and the dotted lines denote the functions $X_\theta$
and $X_\phi$, respectively, and the radiative model with $\dot M/\dot M_{Edd}=0.02$
has been used for the mode computation.
The amplitude normalization is given by $\max |iT_{l^\prime}|=1$ at the surface.
}
\end{figure}
\begin{figure}
\epsscale{.5}
\plotone{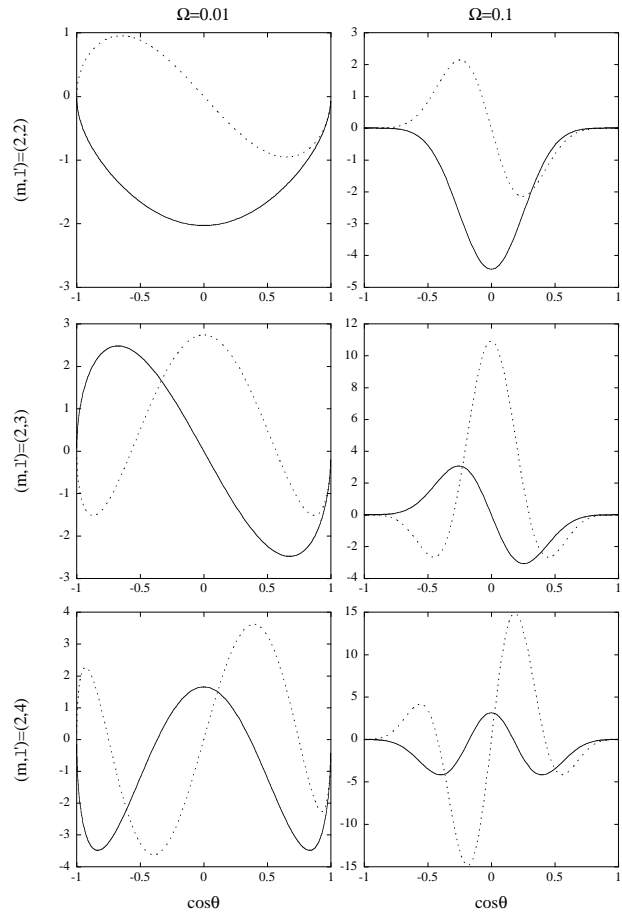}
\caption{Same as Figure 9, but for $m=2$.
}
\end{figure}

\begin{figure}
\epsscale{.5}
\plottwo{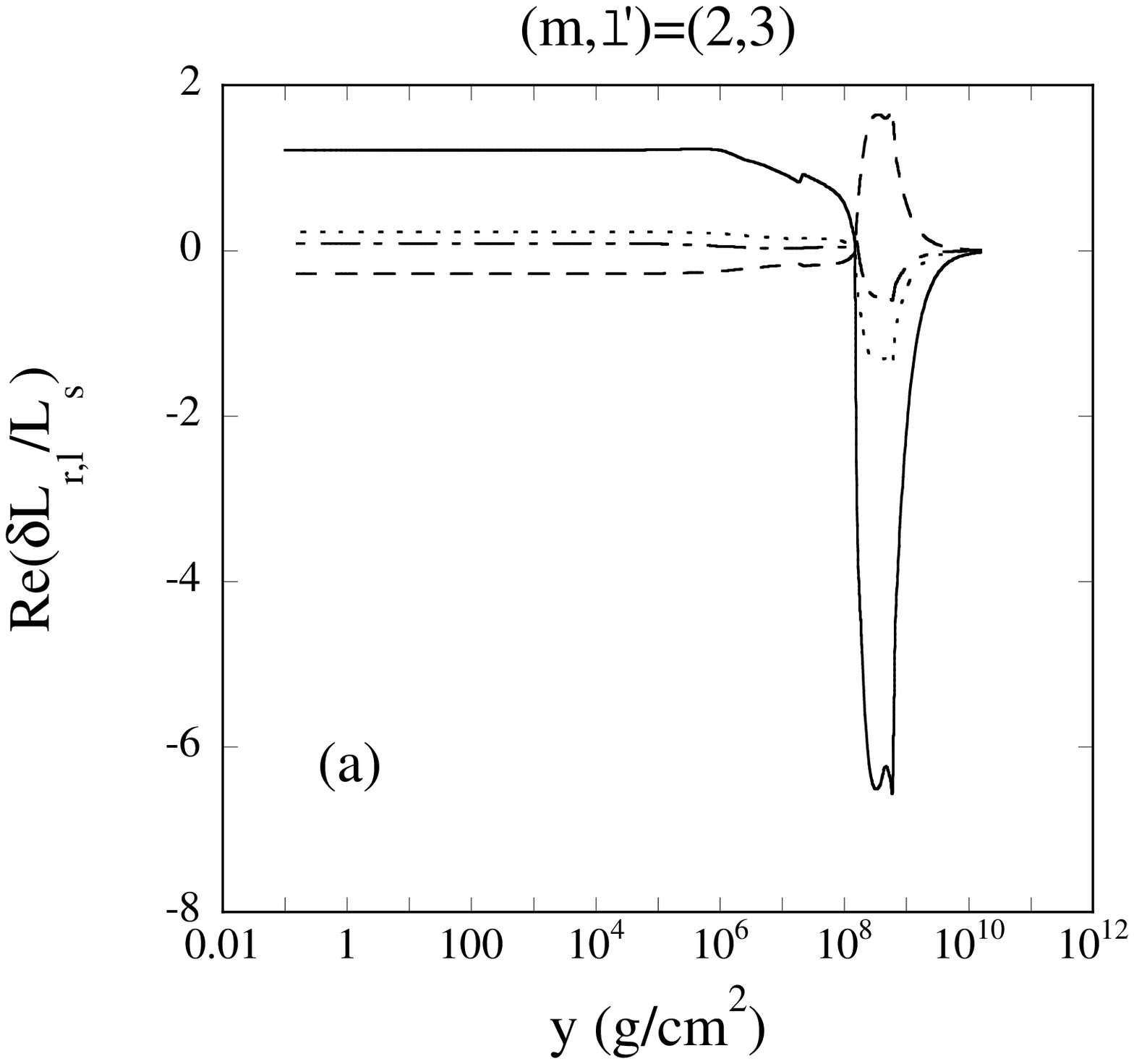}{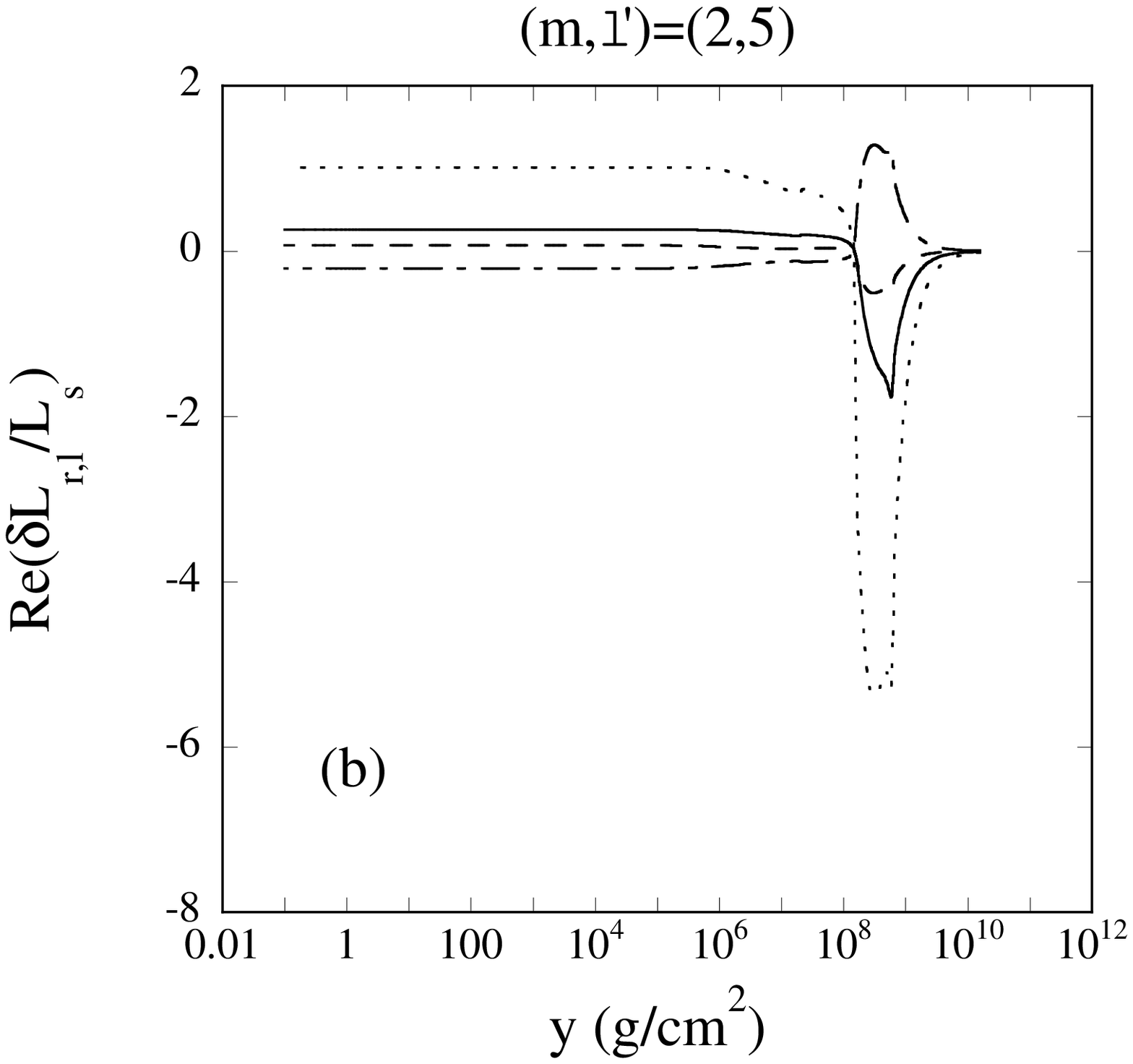}
\caption{Expansion coefficients $\Re(\delta L_{r,l}/L_s)$ of the
fundamental $r$-modes at $\bar\Omega=0.1$ are
given versus the column depth $y$ for the convective model
with $\dot M/\dot M_{Edd}=0.1$, where the solid, dotted, dashed, and dash-dotted lines
are for the coefficients associated with $l=|m|,~|m|+2,~|m|+4$, and $|m|+6$,
respectively.
Panel (a) is for the $r$-mode of $(m,l^\prime)=(2,3)$, and panel (b) is for
the $r$-mode of $(m,l^\prime)=(2,5)$.
The amplitude normalization is given by $\max |iT_{l^\prime}|=1$ at the surface.
}
\end{figure}
\begin{figure}
\epsscale{.5}
\plotone{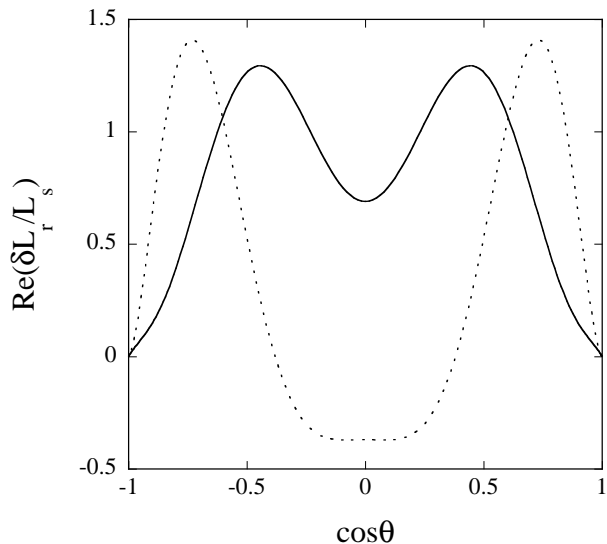}
\caption{
$\Re(\delta L_r/L_s)$ at the surface is plotted as a function of $\cos\theta$
for the fundamental $r$-modes of 
$(m,l^\prime)=(2,3)$ (solid line) and of $(m,l^\prime)=(2,5)$ (dotted line) 
at $\bar\Omega=0.1$ for the convective model
with $\dot M/\dot M_{Edd}=0.1$.
The amplitude normalization is given by $\max |iT_{l^\prime}|=1$ at the surface.
}
\end{figure}

\end{document}